%% file: last-pebble.tex
\documentclass[runningheads,envcountsect,envcountsame]{llncs}
\usepackage[T1]{fontenc}
\usepackage{graphicx}
\input{globals/imports.tex}

\input{globals/macros.tex}

\begin{document}
\title{Pebble minimization: the last theorems}
%
%
\author{Gaëtan Douéneau-Tabot
}
\authorrunning{G. Douéneau-Tabot}
%
\institute{Université Paris Cité, CNRS, IRIF, F-75013, Paris, France\\
\and
Direction générale de l'armement - Ingénierie des projets, Paris, France
\email{doueneau@irif.fr}}
\maketitle              
\begin{abstract} Pebble transducers are nested two-way
transducers which can drop marks (named ``pebbles'') on
their input word. Such machines can compute functions whose
output size is polynomial in the size of their input.
They can be seen as simple recursive programs
whose recursion height is bounded.
A natural problem is, given a pebble transducer,
to compute an equivalent pebble
transducer with minimal recursion height.
This problem has been open since the introduction of the model.

In this paper, we study two restrictions of pebble transducers,  
that cannot see the marks (``blind pebble transducers'' introduced by
Nguy{\^{e}}n et al.),
or that can only see the last mark dropped
(``last pebble transducers'' introduced by Engelfriet et al.). For both models,
we provide an effective algorithm for minimizing the recursion height.
The key property used in both cases is that a function
whose output size is linear (resp. quadratic, cubic, etc.)
can always be computed by a machine whose recursion
height is $1$ (resp. $2$, $3$, etc.). We finally show that this key property
fails as soon as we consider machines that can see more than one mark.

\end{abstract}

\section{Introduction}

Transducers are finite-state machines obtained by adding outputs
to finite automata. They are very useful in a lot of areas like coding, computer
arithmetic, language processing or program analysis, and more
generally in data stream processing. In this paper, we consider
deterministic transducers which compute
functions from finite words to finite words. In particular,
a \reintro{deterministic two-way transducer} is a two-way automaton
with outputs. This model describes the class of \reintro{regular functions}, which is often considered
as one of the functional counterparts of regular languages. It has been
intensively studied for its properties such as
closure under composition~\cite{chytil1977serial}, equivalence
with logical transductions~\cite{engelfriet2001mso} or
regular expressions~\cite{dave2018regular}, decidable
equivalence problem~\cite{gurari1982equivalence}, etc.

\subsubsection{Pebble transducers and polyregular functions.} 
\kl{Two-way transducers} can only describe functions
whose output size is at most linear in the input size.
A possible solution to overcome this limitation
is to consider nested \kl{two-way transducers}.
In particular, the model of \intro{$k$-pebble transducer} has been studied
for a long time \cite{engelfriet2007xml}. For $k=1$, a \kl{$1$-pebble transducer}
is just a \kl{two-way transducer}. For $k \ge 2$, a \kl{$k$-pebble transducer}
is a \kl{two-way transducer} that, when 
on any position $i$ of its input word, can call
a \kl{$(k{-}1)$-pebble transducer}. The latter takes as input
the original input where position $i$ is marked by a ``pebble''.
The main \kl{two-way transducer} then outputs the concatenation
of all the outputs produced along its calls.
The intuitive behavior of a \kl{$3$-pebble transducer} is
depicted in \cref{fig:pebble}. It can be seen
as a recursive program whose recursion stack has  height $3$.
The class of functions computed
by \kl{pebble transducers} is known as \intro{polyregular functions}.
It has been intensively studied due to its properties such as
closure under composition~\cite{engelfriet2015two},
equivalence with logical interpretations~\cite{bojanczyk2019string}, etc.

\vspace*{0.4cm}

\input{parts/fig-pebble-transducer.tex}

\vspace*{-0.5cm}

\subsubsection{Optimization of pebble transducers.} Given a \kl{$k$-pebble transducer}
computing a function $f$, a very natural problem
is to compute the least possible $1 \le \ell \le k$ such that $f$
can be computed by an \kl{$\ell$-pebble transducer}. Furthermore,
we can be interested in effectively building an
\kl{$\ell$-pebble transducer} for $f$. Both questions are open,
but they are meaningful since they ask whether we can
optimize the recursion height (i.e. the running time) of a program.

\AP
It is easy to observe that if $f$ is computed by
a \kl{$k$-pebble transducer}, then $|f(u)| = \bigO(|u|^k)$.
It was first claimed in a LICS 2020 paper that the minimal
recursion height $\ell$ of $f$ (i.e. the least possible $\ell$ such that 
$f$ can be computed by an \kl{$\ell$-pebble transducer}) was exactly
the least possible $\ell$ such that $|f(u)| = \bigO(|u|^{\ell})$. However,
Boja\'nczyk recently disproved this statement in  \cite[Theorem 6.3]{bojanczyk2022transducers}:
the function \label{page:isq}$\intro*\isqK : u_1 \# \cdots \# u_n 
\mapsto (u_1\#)^n \cdots (u_n\#)^n$ can be computed by 
a \kl{$3$-pebble transducer} and is such that
$|\isqK(u)| = \bigO(|u|^2)$, but it cannot be computed by
a \kl{$2$-pebble transducer}. Other counterexamples
were given in \cite{nguyen2023revisiting} using different proof techniques.
Therefore, computing the
minimal recursion height of $f$ is believed to be hard,
since this value not only depends on the output size of
$f$, but also on the word combinatorics of this output.

\subsubsection{Optimization of blind pebble transducers.}  \AP
A subclass of \kl{pebble transducers}, named 
\reintro{blind pebble transducers}, was recently
introduced in~\cite{nous2020comparison}.
A \kl{blind $k$-pebble transducer} is somehow
a \kl{$k$-pebble transducer}, with the difference
that the positions are no longer marked when
making recursive calls. The behavior of
a \kl{blind $3$-pebble transducer} is depicted in \cref{fig:3-blind-transducer}.
The  class of functions computed by \kl{blind pebble transducers}
is strictly included in \kl{polyregular functions}
\cite{doueneau2022hiding,nous2020comparison}.
The main result of \cite{nous2020comparison} shows that 
for \kl{blind pebble transducers}, the minimal recursion height
for computing a function only depends on the growth
of its output. More precisely, if  $f$ is computed
by a \kl{blind $k$-pebble transducer}, then
the least possible $1 \le \ell \le k$ such that $f$ can be computed by
an \kl{blind $\ell$-pebble transducer} is the least possible $\ell$ such that
 $|f(u)| = \bigO(|u|^{\ell})$.

\input{parts/fig-blind-transducer.tex}

\vspace*{-0.7cm}

\subsubsection{Contributions.} \AP In this paper, we first
give a new proof of the connection between minimal recursion height
and growth of the output for \kl{blind pebble transducers}. Furthermore,
our proof provides an algorithm that, given a function computed
by a \kl{blind $k$-pebble transducer}, builds
a \kl{blind $\ell$-pebble transducer} which computes it,
for the least possible $1 \le \ell \le k$.
This effective result is not claimed in \cite{nous2020comparison},
and our proof techniques significantly differ  from  
theirs. Indeed, we make a heavy use of \reintro{factorization forests},
which have already been used as a powerful tool
in the study of \kl{pebble transducers}
\cite{bojanczyk2022growth,doueneau2021pebble,doueneau2022hiding}.

\AP
Secondly, the main contribution of this paper is to show that the (effective) connection
between minimal recursion height and growth of the output
also holds for the class of \reintro{last pebble transducers}
(introduced in \cite{engelfriet2007xml}).
Intuitively, a \kl{last $k$-pebble transducer} is a 
\kl{$k$-pebble transducer} where a called submachine 
can only see the position of its call, but not the
full stack of the former positions. The behavior of
a \kl{last $3$-pebble transducer}
is depicted in \cref{fig:3-last-transducer}. Observe that
a \kl{blind $k$-pebble transducer} is a restricted 
version of a \kl{last $k$-pebble transducer}.
Formally, we show that if $f$ is computed by a
\kl{last $k$-pebble transducer}, then the least possible $\ell$ such that 
$f$ can be computed by a \kl{last $\ell$-pebble transducer}
is the least possible $\ell$ such that $|f(u)| = \bigO(|u|^{\ell})$.
Furthermore, our proof gives an algorithm that
effectively builds a \kl{last $\ell$-pebble transducer}
computing $f$.

\input{parts/fig-last-transducer.tex}

\vspace*{-0.5cm}

\AP
As a third theorem, we show that our
result for \kl{last pebble transducers} is tight,
in the sense that the connection between minimal recursion height
and growth of the output does not hold for more powerful models.
More precisely, we define the model of
\reintro{last-last $k$-pebble transducers},
which extends \kl{last $k$-pebble transducers}
by allowing them to see the two last positions of the calls
(and not only the last one).
We show that for all $k \ge 1$, there exists
a function $f$ such that $|f(u)| = \bigO(|u|^2)$
and that is computed by a 
\kl{last-last $(2k{+}1)$-pebble transducer},
but cannot be computed by a \kl{last-last $2k$-pebble transducer}.
The proof of this result relies on a counterexample presented
by Boja{\'{n}}czyk in \cite{bojanczyk2022growth}.
\vspace*{-0.25cm}

\subsubsection{Outline.} \AP We introduce
\kl{two-way transducers} in \cref{sec:two-way}.
In \cref{sec:blind-last} we describe \kl{blind pebble transducers}
and \kl{last pebble transducers}. We also state our main results
that connect the minimal recursion height of a function
to the growth of its output. Their proof goes over
\cref{sec:forests,sec:blind-proof,sec:last-proof}.
In \cref{sec:last-last}, we finally show that these results
cannot be extended to two visible marks.

\section{Preliminaries on two-way transducers}

\label{sec:two-way}

\AP
Capital letters $A,B$ denote alphabets,
i.e. finite sets of letters.
The empty word is denoted by $\movi$. 
If $u \in A^*$, let $|u| \in \Nat$ be its length,
and for $1 \le i \le |u|$ let $u[i]$ be its $i$-th letter.
If $ i \le j$, we let $u[i{:}j]$ be $u[i]u[i{+}1] \cdots u[j]$
(empty if $j <i$). 
If $a \in A$, let $|u|_a$ be the number of letters $a$ occurring in $u$. 
We assume that the reader is familiar with the basics of automata theory,
in particular two-way automata and monoid morphisms.
The type of total (resp. partial, i.e. possibly undefined on some inputs)
functions is denoted $S \fonc T$ (resp. $S \parfonc T$).

\AP
The machines described in this paper are always
\intro{deterministic}.

\begin{definition}
\label{def:det-two-way}
	\AP
	A \intro{two-way transducer}
	$\trans = (A,B,Q,q_0,F,\delta,\lambda)$ consists of:

	\begin{itemize}
	\item an input alphabet $A$ and an output alphabet $B$;
	\item a finite set of states $Q$
	with $q_0 \in Q$ initial and $F \subseteq Q$ final;
	\item a transition function $\delta: Q \times (A \uplus \{{\lmark}, {\rmark}\})
	\parfonc Q \times \{\lmove, \rmove\}$;
	\item an output function $\lambda: Q \times (A \uplus \{{\lmark}, {\rmark}\}) \parfonc B^*$ with same domain as $\delta$.

	\end{itemize}

\end{definition}

\AP
The semantics of a \kl{two-way transducer} $\trans$ is defined
as follows. When given as input a word $u \in A^*$, 
$\trans$ disposes of a read-only input tape containing
${\lmark} u {\rmark}$. The marks ${\lmark}$ and ${\rmark}$ are used to
detect the borders of the tape, by convention we denote them by positions 
$0$ and $|u|{+}1$ of $u$.
Formally, a configuration over  ${\lmark} u {\rmark}$ is a tuple $(q,i)$
where $q \in Q$ is the current state and $0 \le i \le |u|{+}1$ is the
position of the reading head. The transition relation $\runs{}$
is defined as follows. Given a configuration $(q,i)$,
let $(q',\star):= \delta(q,u[i])$. Then $(q, i) \runs{} (q', i')$ whenever
either $\star = \lmove$ and  $i' = i{-}1$ (move left), or $\star = \rmove$
and $i' = i{+}1$ (move right), with $0 \le i' \le |u|{+}1$. A run is a
sequence of configurations $(q_1,i_1) \runs{} \cdots \runs{} (q_n,i_n)$.
Accepting runs are those that begin in $(q_0, 0)$ and end in a configuration
of the form $(q, |u|{+}1)$ with $q \in F$ (and never visit such a configuration before).

\AP
The partial function $f: A^* \parfonc B^*$ computed by
the  \kl{two-way transducer}  $\trans$ is defined as follows:
for $u \in A^*$, if there exists an accepting run on ${\lmark} u {\rmark}$,
then it is unique, and $f(u)$ is defined as
$\lambda(q_1,(\lmark u \rmark)[i_1]) \cdots \lambda(q_n,(\lmark u \rmark)[i_n]) \in B^*$.
The class of functions computed by
\kl{two-way transducers} is called
\intro{regular functions}.

\begin{example}
Let $\wt{u}$ be the mirror image of $u \in A^*$. Let $\# \not \in A$
be a fresh symbol. The function $\intro*\mapRevK : u_1 \# \cdots \# u_n \mapsto
\wt{u_1} \# \cdots \# \wt{u_n}$
can be computed by a \kl{two-way transducer},
that reads each factor $u_j$ from right to left.
\end{example}

\AP
It is well-known that 
the domain of a \kl{regular function} is always a \intro{regular language}
(see e.g. \cite{shepherdson1959reduction}).
From now on, we assume without losing generalities 
that our \kl{two-way transducers} only compute total functions
(in other words, they have exactly one accepting run on each $\lmark u \rmark$).
Furthermore, we assume that  
$\lambda(q, \lmark) = \lambda(q, \rmark) = \movi$ for all $q \in Q$
(we only lose generality for the image of $\movi$).

\AP
In the rest of this section, $\trans$ denotes a \kl{two-way transducer}
with input alphabet $A$, output alphabet $B$ and output function $\lambda$.
Now, we define the \reintro{crossing sequence} in a position $1 \le i \le |u|$
of input $\lmark u \rmark$.
Intuitively, it regroups the states of the accepting run
which are visited in this position.

\begin{definition}
Let $u \in A^*$ and $1 \le i \le |u|$ .
Let $(q_1,i_1) \runs{} \cdots \runs{} (q_n,i_n)$ be the accepting
run of $\trans$ on $\lmark u \rmark$. The \intro{crossing sequence}
of $\trans$ in $i$, denoted $\intro*\crossK{\trans}{u}(i)$,
is defined as the sequence $(q_j)_{1 \le j \le n \teq{ and } i_j = i}$.
\end{definition}
\AP
If $\mu : A^* \fonc \Mono$ is a monoid morphism,
we say that any $m,m' \in \Mono$ and $a \in A$ define a \intro{$\mu$-context}
that we denote by $m \cro{a} m'$. It is well-known that the
\kl{crossing sequence} in a position of the input only depends
on the \kl{context} of this position, for a well-chosen monoid,
as claimed in \cref{prop:transition-morphism}
(see e.g. \cite{dave2018regular}).

\begin{proposition}
\label{prop:transition-morphism}
One can build a finite monoid $\Tb$ and a
monoid morphism $\mu : A^* \fonc \Tb$, called the
 \intro(TDT-F){transition morphism} of $\trans$, such that
for all $u \in A^*$ and $1 \le i \le |u|$,
$\crossK{\trans}{u}(i)$ only depends on
$\mu(u[1{:}i{-}1]), u[i]$ and $\mu(u[i{+}1{:}|u|])$.\\
Thus we denote it $\reintro*\crossK{\trans}{}(\mu(u[1{:}i{-}1]) \cro{u[i]} \mu(u[i{+}1{:}|u|])$.
\end{proposition}

Finally,  let us define ``the output produced below position $i$''.

\begin{definition}
Let $u \in A^*$ and $1 \le i \le |u|$ and
$q_1 \cdots q_n \defined \crossK{\trans}{u}(i)$. We define
the \intro{production} of $\trans$ in $i$, denoted $\intro*\valK{\trans}{u}(i)$,
as $\lambda(q_1, u[i]) \cdots \lambda(q_n, u[i])$.
\end{definition}
By \cref{prop:transition-morphism}, it also makes sense to define 
$\reintro*\valK{\trans}{}(m \cro{a} m') \in B^*$ to be $\valK{\trans}{u}(i)$ whenever
$m = \mu(u[1{:}i{-}1])$, $m' = \mu(u[i{+}1{:}|u|])$ and $a = u[i]$.

\section{Blind and last pebble transducers}

\label{sec:blind-last}

\AP
Now, we are ready to define formally the models of
\kl{blind pebble transducers} and \kl{last pebble transducers}.
Intuitively, they correspond to \kl{two-way transducers}
which make a tree of recursive calls to other \kl{two-way transducers}.

\begin{definition}[Blind pebble transducer \cite{nous2020comparison}]
\label{def:blind-pebble} \AP
For $k \ge 1$, a \intro{\mbox{blind $k$-pebble} transducer} with input alphabet $A$
and output alphabet $B$ is:
\begin{itemize}
\item if $k =1$, a \kl{two-way transducer} with input alphabet $A$
and output $B$;
\item if $k \ge 2$, a tree $\trans\tree{\btrans_1, \cdots ,\btrans_p}$ where
the subtrees $\btrans_1, \dots, \btrans_p$ are \kl{blind $(k{-}1)$-pebble transducers}
with input $A$ and output $B$; and the root label $\trans$ is a \kl{two-way transducer}
with input $A$ and output alphabet $\{\btrans_1, \dots, \btrans_p\}$.
\end{itemize}
\end{definition}
\AP
The (total) function $f:A^* \fonc B^*$
computed by the \kl{blind $k$-pebble transducer} 
of \cref{def:blind-pebble} is built in a recursive fashion, as
follows:
\begin{itemize}
\item for $k=1$, $f$ is the function computed by the \kl{two-way transducer};
\item for $k\ge 2$, let $u \in A^*$ and  $(q_1,i_1) \runs{} \cdots \runs{} (q_n,i_n)$ be the
accepting run of $\trans = (A,B,Q,q_0,F,\delta,\lambda)$ on $\lmark u \rmark$.
For all $1 \le j \le n$, let $f_j : A^* \fonc B^*$ be the concatenation of the functions
recursively computed by the sequence $\lambda(q_j,(\lmark u\rmark)[i_j]) \in \{\btrans_1, \dots, \btrans_p\}^*$.
Then $f(u) \defined f_1 (u) \cdots f_n(u)$.
\end{itemize}
The behavior of a \kl{blind $3$-pebble transducer} is depicted
in \cref{fig:3-blind-transducer}.

\begin{example}  \AP \label{ex:ulsq}
The function $\intro*\ulsq : A^* \fonc A^* \uplus \{\#\}, u \mapsto (u\#)^{|u|}$
can be computed by a \kl{blind $2$-pebble transducer}.
This machine has shape $\trans \tree{\trans'}$:
$\trans$ calls $\trans'$ on each position
$1 \le i \le |u|$ of its input $u$, and $\trans'$ outputs $u\#$.
\end{example}
\AP
The class of functions computed by a 
\kl{blind $k$-pebble transducer} for some $k \ge 1$
is called \intro{polyblind functions} \cite{doueneau2022hiding}.
They form a strict subclass of \kl{polyregular functions}
\cite{doueneau2021pebble,doueneau2022hiding,nous2020comparison} which is closed
under composition \cite[Theorem 6.1]{nous2020comparison}.

\AP
Now, let us define \kl{last pebble transducers}. They corresponds to
\kl{blind pebble transducers} enhanced with the ability to
mark the current position of the input when doing a recursive call.
Formally, this position is underlined and we define
$\intro*\omarK{u}{i} \defined u[1] \cdots u[i{-}1] \marq{u[i]} u[i{+}1] \cdots u[|u|]$
for $u \in A^*$ and $1 \le i \le |u|$.

\begin{definition}[Last pebble transducer \cite{engelfriet2007xml}]
\label{def:last-pebble} \AP
For $k \ge 1$, a \intro{last $k$-pebble transducer}
with input alphabet $A$
and output alphabet $B$ is:
\begin{itemize}
\item if $k =1$, a \kl{two-way transducer} with input alphabet $A \uplus \marq{A}$
and output $B$;
\item if $k \ge 2$, a tree $\trans\tree{\ltrans_1, \cdots ,\ltrans_p}$ where
the subtrees $\ltrans_1, \dots, \ltrans_p$ are \kl{last \linebreak $(k{-}1)$-pebble transducers}
with input $A$ and output $B$; 
and the root label $\trans$ is a \kl{two-way transducer} with input $A \uplus \marq{A}$
and output alphabet $\{\ltrans_1, \dots, \ltrans_p\}$.
\end{itemize}
\end{definition}
\AP
The (total) function  $f:(A \uplus \marq{A})^* \fonc B^*$ computed by
the \kl{last $k$-pebble transducer} of \cref{def:last-pebble} is defined in a recursive
fashion, as follows:
\begin{itemize}
\item for $k=1$, $f$ is the function computed by
the \kl{two-way transducer};
\item for $k\ge 2$, let $u \in A^*$ and  $(q_1,i_1) \runs{} \cdots \runs{} (q_n,i_n)$ be the
accepting run of $\trans= (A \uplus \marq{A},B,Q,q_0,F,\delta,\lambda)$ on $\lmark u \rmark$.
For all $1 \le j \le n$, let $f_j : A^* \fonc B^*$ be the concatenation of the functions
recursively computed by $\lambda(q_j,(\lmark u \rmark)[i_j]) \in \{\ltrans_1, \dots, \ltrans_p\}^*$.
Let   $\tau : (A \uplus \marq{A})^* \fonc A^*$ be
the morphism which erases the underlining (i.e.
$\tau(\marq{a})=a$), then
$f(u) \defined f_1(\omarK{\tau(u)}{i_1}) \cdots f_n(\omarK{\tau(u)}{i_n})$.
\end{itemize}
The behavior of a \kl{last $3$-pebble transducer} is depicted
in \cref{fig:3-last-transducer}. Observe that our definition
builds a function of type $(A \uplus \marq{A})^* \fonc B^*$,
but we shall in fact consider its restriction to $A^*$ (the marks
are only used within the induction step).

\begin{example}[\cite{bojanczyk2018polyregular}]
\label{ex:square}
The function $\intro*\sqK :  u \mapsto (\omarK{u}{1})\# \cdots
 (\omarK{u}{|u|})\#$ can be computed by a \kl{last $2$-pebble transducer},
 which successively marks and makes recursive calls in positions $1, 2$, etc.
However this function is not \kl{polyblind} \cite{nous2020comparison}.
\end{example}

\AP
We are ready to state our main result. Its proof goes
over \cref{sec:forests,sec:blind-proof,sec:last-proof}.

\begin{theorem}[Minimization of the recursion height]
\label{the:minimize-main}
Let $1 \le \ell \le k$. Let $f : A^* \fonc B^*$ be
computed by a  \kl{blind $k$-pebble transducer} (resp.
by a \kl{last $k$-pebble transducer}).
Then $f$ can be computed by a \kl{blind $\ell$-pebble transducer}
 (resp. by a \kl{last $\ell$-pebble transducer})
if and only if $|f(u)| = \bigO(|u|^{\ell})$.\\
This property is decidable and the construction
is effective.
\end{theorem}

\AP
As an easy consequence, the class of functions
computed by \kl{last pebble transducers} form a strict subclass
of the \kl{polyregular functions} (because \cref{the:minimize-main}
does not hold for the full model of \kl{pebble transducers} \cite[Theorem 6.3]{bojanczyk2022transducers})
and therefore it is not closed under composition (because
any \kl{polyregular function} can be obtained
as a composition of \kl{regular functions} and $\sqK$s
\cite{bojanczyk2018polyregular}).

\AP
Even if a (non-effective) \cref{the:minimize-main} was already known for \kl{blind pebble transducers}
\cite[Theorem 7.1]{nous2020comparison}, we shall first present
our proof of this case. Indeed, it is a new proof (relying on
\kl{factorization forests}) which is simpler than the original one.
Furthermore, understanding the techniques used is a key step
for understanding the proof for \kl{last pebble transducers}
presented afterwards.

\section{Factorization forests}

\label{sec:forests}

In this section, we introduce the key tool of \kl{factorization forests}.
Given a monoid morphism $\mu:A^* \rightarrow \Mono$
and $u \in A^*$, a \kl{$\mu$-factorization forest}
of $u$ is an unranked tree structure defined as follows.
We use the brackets $\tree{\cdots}$ to build a tree.

\begin{definition}[{Factorization forest} \cite{simon1990factorization}]%
\label{def:facto} \AP
Given a morphism $\mu: A^* \rightarrow \Mono$ and $u\in A^*$, we say that
$\forest$ is a  \intro{$\mu$-forest} of $u$ if:
\begin{itemize}
\item either $u = \movi$ and $\forest = \movi$; or $u = \tree{a}  \in A $ and $\forest = a$;
\item or $\forest = \tree{\forest_1, \cdots ,\forest_n}$,
$u = u_1 \cdots u_n$, for all $1 \le i \le n$,
$\forest_i$ is a \kl{$\mu$-forest} of $u_i \in A^+$,
and if $n \ge 3$ then $\mu(u) = \mu(u_1) = \dots = \mu(u_n)$ is idempotent.
\end{itemize}
\end{definition}
\AP
We use the standard tree vocabulary of height,
child, sibling, descendant and ancestor
(a node being itself one of its ancestors/descendants), etc.
We denote by $\intro*\NodK{\forest}$ the set of nodes of $\forest$.
In order to simplify the statements, we identify a node $\nod \in \NodK{\forest}$ with
the subtree rooted in this node. Thus $\NodK{\forest}$ can also
be seen as the set of subtrees of $\forest$, and $\forest \in \NodK{\forest}$.
We say that a node is \intro(NOD){idempotent} if it has
at least $3$ children. We denote by $\intro*\FacK{\mu}{}(u)$ (resp. $\reintro*\FacK{\mu}{d}(u)$)
the set of \kl{$\mu$-forests} of $u \in A^*$ (resp. 
\kl{$\mu$-forests} of $u \in A^*$ of height at most $d$).
We write $\reintro*\FacK{\mu}{}$ and $\reintro*\FacK{\mu}{d}$
of all \kl{forests} (of any word).

A $\mu$-forest of $u \in A^*$ can also be seen as
``the word $u$ with brackets'' in \cref{def:facto}.
Therefore $\FacK{\mu}{}$ can be seen as a language over
 $\apar{A} \defined A \uplus \{\lefttree,\righttree\}$. In this setting,
 it is well-known that \kl{$\mu$-forests} of bounded height can
 effectively be computed by a \intro{rational function}, i.e. a
 particular case of \kl{regular function} that can be computed
 by a non-deterministic one-way transducer (see e.g. \cite{doueneau2021pebble}).

\begin{theorem}[Simon \cite{simon1990factorization,colcombet2011green}]%
	\label{the:simon}
	\AP
	Given a morphism $\mu : A^* \to \Mono$ into
	a finite monoid $\Mono$, one can effectively build a \kl{rational function}
	$\intro*\forK{\mu} : A^* \to (\apar{A})^*$
	such that for all $u \in A^*$, $\forK{\mu}(u) \in \FacK{\mu}{3|\Mono|}(u)$.
\end{theorem}

\AP
Building \kl{$\mu$-forests} of bounded height is especially useful
for us, since it enables to decompose
any word in a somehow bounded way. This decomposition
will be guided by the following definitions, that have 
been introduced in \cite{doueneau2021pebble,doueneau2022hiding}.
First, we define \kl{iterable nodes} as the middle
children of \kl{idempotent nodes}.

\begin{definition}%
\label{def:iterables}
\AP
Let $\forest \in \FacK{\mu}{}(u)$. Its \intro{iterable nodes},
denoted $\intro*\IteK{\forest}$, are:
\begin{itemize}
\item if $\forest = \tree{a}\in A$ or $\forest = \movi$,
then $\IteK{\forest} \defined \vide$;
\item otherwise if $\forest = \tree{\forest_1, \cdots ,\forest_n}$, then:
\begin{equalign*}
\IteK{\forest} \defined \{\forest_i: 2 \le i \le n{-}1\} \cup  \bigcup_{1 \le i \le n} \IteK{\forest_i}.
\end{equalign*}
\end{itemize}
\end{definition}
\AP
Now, we define the notion of \kl{skeleton} of a node $\nod$,
which contains all the descendants
of $\nod$ except those which are \kl(NOD){iterable}.

\begin{definition}[Skeleton, frontier]%
\label{def:skeleton}
\AP
Let $\forest \in \FacK{\mu}{}(u)$, $\nod \in \NodK{\forest}$, we define
the \intro{skeleton}  of $\nod$, denoted $\intro*\SkeK{\forest}(\nod)$, by:
\begin{itemize}
\item if $\nod = \tree{a}\in A$ is a leaf, then $\SkeK{\forest}(\nod) \defined \{ \nod \}$;
\item otherwise if $\nod = \tree{\forest_1, \cdots, \forest_n}$, then
$\SkeK{\forest}(\nod) \defined \{\nod\} \cup \SkeK{\forest}(\forest_1) \cup \SkeK{\forest}(\forest_n)$.
\end{itemize}
The \intro{frontier} of $\nod$ is the set $\intro*\FroK{\forest}(\nod) \subseteq [1{:}|u|]$
containing the positions of $u$ which belong to $\SkeK{\forest}(\nod)$
(when seen as leaves of the \kl{$\mu$-forest} $\forest$ over $u$).
\end{definition}

\begin{example}
Let $\Mono \defined (\{-1,1,0\}, \times)$ and
$\mu: \Mono^* \fonc \Mono$ the product.
A \kl{$\mu$-forest} $\forest$ of the
word $(-1)(-1)0(-1)000000$ is depicted
in Figure~\ref{fig:ex:skel}. Double lines
denote \kl{idempotent nodes}. The set of
blue nodes is the \kl{skeleton} of the topmost
blue node.
\end{example}

\input{parts/fig-factorization-forest.tex}

\AP
It is easy to observe that for $\forest \in \FacK{\mu}{d}(u)$, the size
of a \kl{skeleton}, or of a \kl{frontier}, is bounded independently
from $\forest$. Furthermore,
the set of \kl{skeletons} $\{ \SkeK{\forest}(\nod) : \nod \in \IteK{\forest} \cup \{\forest\} \}$
is a partition of $\NodK{\forest}$ \cite[Lemma 33]{doueneau2021pebble}.
As a consequence, the set of \kl{frontiers} 
$ \{\FroK{\forest}(\nod): \nod \in \IteK{\forest} \cup \{\forest\}\}$
is a partition of $[1{:} |u|]$.
Given a position $1 \le i \le |u|$, we can
thus define the \intro{origin} of $i$ in $\forest$,
denoted $\intro*\originK{\forest}(i)$, as the 
unique $\nod \in \IteK{\forest} \cup \{\forest\}$
such that  $i \in \FroK{\forest}{}(\nod)$.

\begin{definition}[Observation]{}
	Let $\forest \in \FacK{\mu}{}$ and $\nod, \nod' \in \NodK{\forest}$.
	We say that $\nod \in  \NodK{\forest}$  \intro(NOD){observes} $\nod' \in \NodK{\forest}$
	if either $\nod'$ is an ancestor of $\nod$, or $\nod'$ is the immediate
	right or left sibling of an ancestor of $\nod$.
\end{definition}

\input{parts/fig-observable.tex}

\AP
The intuition behind the notion of \kl(NOD){observation}
(which is \emph{not} symmetrical) is depicted in \cref{fig:observation}.
Note that in a forest of bounded height,
the number of nodes that some $\nod$ \kl(NOD){observes} is bounded.
This will be a key argument in the following.
We say that $\nod$ and $\nod'$ are \intro(NOD){dependent} if
either $\nod$ \kl(NOD){observes} $\nod'$ or the converse.
Given $\forest$, we can translate these notions
to the positions of $u$: we say that $i$ \intro(POS){observes}
(resp. \intro(POS){depends} on) $i'$ if
$\originK{\forest}(i)$ \kl(NOD){observes} (resp.
\kl(NOD){depends} on) $\originK{\forest}(i')$.

\section{Height minimization of blind pebble transducers}

\label{sec:blind-proof}

\AP
In this section, we show \cref{the:minimize-main} for \kl{blind pebble transducers}.
We say that a \kl{two-way transducer} $\trans$ is a  \intro(BLIND){submachine} of a
\kl{blind pebble transducer} $\btrans$ if $\trans$ labels a node in 
the tree description of $\btrans$. If $\btrans = \trans \tree{\btrans_1, \dots, \btrans_n}$,
we say that the \kl(BLIND){submachine} $\trans$ is the \intro(BLIND){head} of $\btrans$.
We let the \intro(BLIND){transition morphism} 
of $\btrans$ be the cartesian product of all the
 \kl(TDT-F){transition morphisms}
of all the \kl(BLIND){submachines} of $\btrans$.
Observe that it makes sense to consider the \kl{production}
of a \kl(BLIND){submachine} $\trans$ in a  \kl{context} defined
using the \kl(BLIND){transition morphism} of $\btrans$.

\subsection{Pumpability}

\AP
We first give a sufficient condition, named \kl(BLIND){pumpability},
for a \kl{blind $k$-pebble transducer} to compute a
function $f$ such that $|f(u)| \neq \bigO(|u|^{k-1})$. The behavior
of a \kl(BLIND){pumpable} \kl{blind $2$-pebble transducer} is
depicted in \cref{fig:blind-pumpable} over a well-chosen input: it has a factor
in which the \kl(BLIND){head} $\trans_1$ calls a \kl(BLIND){submachine}
$\trans_2$, and a factor in which $\trans_2$ produces a non-empty output.
Furthermore both factors can be iterated without destroying
the runs of these machines (due to idempotents).

\begin{definition}
\label{def:pumpable-blind}
\AP
Let $\btrans$ be a  \kl{blind $k$-pebble transducer}
whose \kl(BLIND){transition morphism} is $\mu : A^* \fonc \Tb$.
We say that the transducer $\btrans$ is \intro(BLIND){pumpable} if there exists:
\begin{itemize}
\item \kl(BLIND){submachines} $\trans_1, \dots, \trans_k$ of $\btrans$,
such that $\trans_1$ is the \kl(BLIND){head} of $\btrans$;
\item  $m_0, \dots, m_k, {\ell}_1, \dots, {\ell_k}, r_1, \dots, r_k \in \mu(A^*)$;
\item $a_1,\dots, a_k \in A$  such that for all $1 \le j \le k$,
$e_j \defined {\ell}_j \mu(a_j) r_j$ is an idempotent;
\item  a permutation $\sigma : \inter{1}{k} \fonc  \inter{1}{k}$;
\end{itemize}
such that if $\Mblock{i}{j}  \defined
m_{i} e_{i + 1} m_{i + 1} \cdots e_{j} m_{j}$
 for all $0 \le i \le j \le k$,  and
 if we define the following \kl{context} for all $1 \le j \le k$:
\begin{equalign*}
\Ccont_j \defined \Mblock{0}{\sigma(j)-1} e_{\sigma(j)} \ell_{\sigma(j)}
\cro{a_{\sigma(j)}} r_{\sigma(j)} e_{\sigma(j)} \Mblock{\sigma(j)}{k}
\end{equalign*}
then for all $1 \le j \le k{-}1$, $|\valK{\trans_j}{}(\Ccont_j)|_{\trans_{j+1}} \neq 0$,
and $\valK{\trans_k}{}(\Ccont_k) \neq \movi$.
\end{definition}

\input{parts/fig-blind-pumpable.tex}

\Cref{lem:growth-blind} follows
by choosing inverse images in $A^*$
for the $m_i$, $\ell_i$ and $r_i$.

\begin{lemma}
\label{lem:growth-blind}
Let $f$  be computed by a
\kl(BLIND){pumpable} \kl{blind $k$-pebble transducer}.
There exists words $v_0, \dots, v_k, u_1, \dots, u_k$
such that $|f(v_0 u_1^{X} \cdots u_k^{X} v_k)| = \bigT(X^k)$.
\end{lemma}

\AP
Now, we use \kl(BLIND){pumpability} as a key ingredient
for showing  \cref{the:minimize-main}, which directly follows
 by induction from the more precise \cref{the:blind-minimize}.

\begin{theorem}[Removing one layer]
\label{the:blind-minimize}
Let $k \ge 2$ and $f : A^* \fonc B^*$ be
computed by a  \kl{blind $k$-pebble transducer} $\btrans$. 
The following are equivalent:
\begin{enumerate}
\item \label{it:bm:growth} $|f(u)| = \bigO(|u|^{k-1})$;
\item \label{it:bm:pumpable} $\btrans$ is not \kl(BLIND){pumpable};
\item \label{it:bm:blind} $f$ can be computed by a \kl{blind $(k{-}1)$-pebble transducer}.
\end{enumerate}
Furthermore, this property is decidable and the construction
is effective.
\end{theorem}

\begin{proof} \Cref{it:bm:blind} $\implies$ \cref{it:bm:growth} is obvious.
\Cref{it:bm:growth} $\implies$ \cref{it:bm:pumpable} is \cref{lem:growth-blind}.
Furthermore, \kl(BLIND){pumpability} can be tested by an enumeration of $\mu(A^*)$ and $A$.
It remains to show \cref{it:bm:pumpable} $\implies$ \cref{it:bm:blind}
(in an effective fashion): this is the purpose of \cref{ssec:minimize-blind}.
\end{proof}

\subsection{Algorithm for removing a recursion layer}

\label{ssec:minimize-blind}

\AP
Let $k \ge 2$ and $\utrans$ be a \kl{blind $k$-pebble transducer} that is not
\kl(LAST){pumpable}, and that computes $f : A^* \fonc B^*$.
We build a \kl{blind $(k{-}1)$-pebble transducer} $\ov{\utrans}$ for $f$.

\AP
Let $\mu : A^* \fonc \Tb$ be the \kl(BLIND){transition morphism}
of $\utrans$. We shall consider that, on input $u \in A^*$,
the \kl(BLIND){submachines} of $\ov{\utrans}$ can in fact use
$\forK{\mu}(u) \subseteq (\apar{A})^*$ as input.
Indeed $\forK{\mu}$ is a \kl{rational function}
(by \cref{the:simon}), hence its information can
be recovered by using a \intro{lookaround}. Informally, the \kl{lookaround} feature
enables a \kl{two-way transducer} to chose its transitions
not only depending on its current state and current letter $u[i]$
in position $1 \le i \le |u|$, but also on a regular property
of the prefix $u[1{:}i{-}1]$ and the suffix $u[i{+1}{:}|u|]$. 
It is well-known that given a \kl{two-way transducer} $\trans$
with \kl{lookarounds}, one can build an equivalent $\trans'$ that
does not have this feature (see e.g.  \cite{hopcroft1967approach,engelfriet2001mso}).
Furthermore, even if the accepting runs of $\trans$ and $\trans'$ may differ,
they produce the same outputs from the same positions (this observation
will be critical for \kl{last pebble transducers}, in order to ensure that
the marked positions of the recursive calls will be preserved).

\AP
Now, we describe the \kl{two-way transducers}
that are the  \kl(BLIND){submachines} of $\ov{\utrans}$.
First, it has  \kl(BLIND){submachines} $\reintro*\normalK{\trans}$
for $\trans$ a  \kl(BLIND){submachine}  of $\utrans$,
which are described in  \cref{algo:blind-minimization-original}.
Intuitively, $\normalK{\trans}$ is just a copy of $\trans$.
It is clear that if $\trans$ is a \kl(BLIND){submachine} of $\utrans$,
then $\normalK{\trans}(u)$ is the concatenation of the outputs produced by (the recursive
calls of) $\trans$ along its accepting run on $\lmark u \rmark$.

\input{parts/algo-blind-minimization-normal.tex}

\AP
$\ov{\utrans}$ also has \kl(BLIND){submachines} $\reintro*\simBlindK{\trans}$
for $\trans$ a  \kl(BLIND){submachine}  of $\utrans$,
which are described in  \cref{algo:blind-minimization-simplify}.
Intuitively, $\simBlindK{\trans}$ simulates $\trans$
while trying to inline recursive calls in its own run.
More precisely, let $u \in A^*$ be the input and $\forest \defined \forK{\mu}(u)$.
If $\trans$ calls $\btrans'$ in $1 \le i \le |u|$ that belongs to the  \kl{frontier}
of the root node  $\forest$ of $\forest$, then $\simBlindK{\trans}$
inlines the behavior of the \kl(LAST){head} of $\btrans'$. Otherwise
it makes a recursive call, except if $\btrans'$ is a leaf of $\utrans$.
Hence if $\trans$ is a \kl(BLIND){submachine} of $\utrans$ which is not a leaf,
$\simBlindK{\trans}(u)$ is the concatenation of the outputs produced by the calls
of $\trans$ along its accepting run.

\input{parts/algo-blind-minimization-simplify.tex}

\AP
Finally, the transducer $\ov{\utrans}$ is
obtained by defining $\simBlindK{\trans}$
to be its  \kl(BLIND){head}, where $\trans$
is the \kl(BLIND){head} of $\utrans$. Furthermore,
we remove the \kl(BLIND){submachines} $\normalK{\trans}$
or $\simBlindK{\trans}$ which are never called.
Observe that $\ov{\utrans}$ indeed computes the function $f$.
Furthermore, we observe that $\ov{\utrans}$ has recursion height (i.e. the number
of nested \textbf{Call} instructions, plus $1$ for the \kl(BLIND){head}) $k{-}1$,\label{sta:k:blind} since each
inlining of \cref{line:fril,line:2-way-blind,line:no-2dt-without-inline} in \cref{algo:blind-minimization-simplify}
removes exactly one recursion layer of $\utrans$.

\AP
It remains to justify that each $\simBlindK{\trans}$ can
be implemented by a \kl{two-way transducer} (i.e. with \kl{lookarounds}
but a bounded memory). We represent
variable $i$  by the current position of the transducer.
Since it has access to $\forest$, the \kl{lookaround}
can be used to check whether $i \in \FroK{\forest}(\forest)$ or not
(since the size of $\FroK{\forest}(\forest)$ is bounded).
It remains to explain how the inlinings are performed:
\begin{itemize}
\item if $i \in \FroK{\forest}(\forest)$, 
the \kl{two-way transducer} inlines
$\normalK{\trans'}$ by executing the same moves
and calls as $\trans'$ does. Once its computation is ended,
it has to go back to position $i$. This is indeed possible
since belonging to $\FroK{\forest}(\forest)$ is a property
that can be detected by using the \kl{lookaround},
hence the machine only needs to remember
 that $i$ was the $\ell$-th position of $\FroK{\forest}(\forest)$ ($\ell$ being bounded);
\item else if $\btrans' = \trans'$ is a \kl{blind $1$-pebble transducer},
we produce the output of $\trans'$ without moving.
This is possible since for all
$i' \not \in \FroK{\forest}(\forest)$, 
$\valK{\trans'}{u}(i') = \movi$ 
(hence the output of $\trans'$ on $u$ is bounded,
and its value can be determined without moving, just by using the \kl{lookaround}).
Indeed, if $\valK{\trans'}{u}(i') \neq \movi$ for such an $i' \not \in \FroK{\forest}(\forest)$
when reaching \cref{line:2-way-blind} 
of  \cref{algo:blind-minimization-simplify}, then
 the conditions of \cref{lem:k-nodes-blind-key}
hold,\label{sta:k:2} which yields a contradiction.
This lemma is the key argument of this proof, relying on
the non-\kl(BLIND){pumpability} of $\utrans$.
\end{itemize}

\begin{lemma}[Key lemma]%
\label{lem:k-nodes-blind-key}%
Let $u \in A^*$ and $\forest \in \FacK{\mu}{}(u)$.
Assume that there exists a sequence $\trans_1, \dots, \trans_k$
of \kl(BLIND){submachines}
of $\utrans$ and a sequence of positions 
$1 \le i_1, \dots, i_k \le |u|$ such that:
\begin{itemize}
\item $\trans_1$ is the \kl(BLIND){head} of $\utrans$;
\item  for all $1 \le j \le k{-}1$, $|\valK{\trans_j}{u}(i_j)|_{\trans_{j+1}} \neq 0$
and $\valK{\trans_k}{u}(i_k) \neq \movi$;
\item for all $1 \le j \le k$, $i_j \not \in \FroK{\forest}(\forest)$
(i.e. $\originK{\forest}(i_j) \in \IteK{\forest}$).
\end{itemize}
Then $\btrans$ is \kl(BLIND){pumpable}.
\end{lemma}

\begin{proof}[idea] We first observe that \kl(BLIND){pumpability}
follows as soon as the nodes $\originK{\forest}(i_j)$ are
pairwise \kl(NOD){independent}. We then
show that this \kl(NOD){independence} condition
can always be obtained, up to duplicating
some \kl(NOD){iterable} subtrees of $\forest$ (and some factors of $u$),
because the behavior of a \kl(BLIND){submachine}
in a \kl{blind pebble transducer} does not depend
on the positions of the above recursive calls.
\end{proof}

\section{Height minimization of last pebble transducers}

\label{sec:last-proof}

\AP
In this section, we show \cref{the:minimize-main} for \kl{last pebble transducers}.
The notions of \intro(LAST){submachine}, \intro(LAST){head}
and \intro(LAST){transition morphism} for
a \kl{last pebble transducer} are defined as in \cref{sec:blind-proof}.
The \kl(LAST){transition morphism}
is now defined over $(A \uplus \marq{A})^*$.

\subsection{Pumpability}

\AP
The sketch of the proof is similar to \cref{sec:blind-proof}.
We first give an equivalent of \kl(BLIND){pumpability}
for \kl{last pebble transducers}. The intuition behind this notion
is depicted in \cref{fig:last-pumpable}. The formal definition is however
more cumbersome, since we need to keep track of the fact
that the calling position is marked.

\begin{definition}%
\label{def:pumpable-last}
\AP
Let $\ltrans$ be a  \kl{last $k$-pebble transducer}
whose \kl(LAST){transition morphism} is $\mu : (A \cup \marq{A})^* \fonc \Tb$.
We say that the transducer $\ltrans$ is \intro(LAST){pumpable} if there exists:
\begin{itemize}
\item \kl(LAST){submachines} $\trans_1, \dots, \trans_k$ of $\ltrans$,
such that $\trans_1$ is the \kl(LAST){head} of $\ltrans$;
\item  $m_0, \dots, m_k, {\ell}_1, \dots, {\ell_k}, r_1, \dots, r_k \in \mu(A^*)$;
\item $a_1,\dots, a_k \in A$  such that for all $1 \le j \le k$,
$e_j \defined {\ell}_j \mu(a_j) r_j$ is idempotent;
\item  a permutation $\sigma : \inter{1}{k} \fonc  \inter{1}{k}$;
\end{itemize}
such that if we let
$\Mblock{i}{j}  \defined
m_{i} e_{i + 1} m_{i + 1} \cdots e_{j} m_{j}$
 for all $0 \le i \le j \le k$,  and  if we define the following \kl{context}:
\begin{equalign*}
\Ccont_1 \defined \Mblock{0}{\sigma(1)-1} e_{\sigma(1)} \ell_{\sigma(1)} \cro{a_{\sigma(1)}} r_{\sigma(1)}
e_{\sigma(1)} \Mblock{\sigma(1)}{k}
\end{equalign*}
and for all $1 \le j \le k{-}1$ the \kl{context}:
\begin{equalign*}
\Ccont_{j+1} & \defined  \Mblock{0}{\sigma(j)-1}
e_{\sigma(j)} \ell_{\sigma(j)} \mu(\marq{a_{\sigma(j)}}) r_{\sigma(j)} e_{\sigma(j)}
\Mblock{\sigma(j)}{\sigma(j+1)-1}&\\
&e_{\sigma(j+1)} \ell_{\sigma(j+1)}  \cro{a_{\sigma(j+1)}} r_{\sigma(j+1)} e_{\sigma(j+1)}
\Mblock{\sigma(j+1)}{k}
&\teq{\normalfont~if $\sigma(j) < \sigma(j+1)$;}\\
\Ccont_{j+1} & \defined  \Mblock{0}{\sigma(j)-1}
e_{\sigma(j+1)} \ell_{\sigma(j+1)} \cro{a_{\sigma(j+1)}} r_{\sigma(j+1)}  e_{\sigma(j+1)} &\\
& \Mblock{\sigma(j+1)}{\sigma(j)-1}  e_{\sigma(j)} \ell_{\sigma(j)}
\mu(\marq{a_{\sigma(j)}}) r_{\sigma(j)} e_{\sigma(j)}
\Mblock{\sigma(j)}{k}
&\teq{\normalfont~otherwise;}\\
\end{equalign*}
then for all $1 \le j \le k{-}1$, $|\valK{\trans_j}{}(\Ccont_j)|_{\trans_{j+1}} \neq 0$,
and $\valK{\trans_k}{}(\Ccont_k) \neq \movi$.
\end{definition}

\input{parts/fig-last-pumpable.tex}

We obtain \cref{lem:growth-last} by
a proof which is similar to that of \cref{lem:growth-blind}. 

\begin{lemma}%
\label{lem:growth-last}%
Let $f$ be computed by a \kl(LAST){pumpable} \kl{last $k$-pebble transducer}.
There exists words $v_0, \dots, v_k, u_1, \dots, u_k$
such that $|f(v_0 u_1^{X} \cdots u_k^{X} v_k)| = \bigT(X^k)$.
\end{lemma}

\begin{theorem}[Removing one layer]
\label{the:last-minimization}
Let $k \ge 2$ and $f : A^* \fonc B^*$ be
computed by a  \kl{last $k$-pebble transducer} $\ltrans$. 
The following are equivalent:
\begin{enumerate}
\item \label{it:lm:growth} $|f(u)| = \bigO(|u|^{k-1})$;
\item \label{it:lm:pumpable} $\ltrans$ is not \kl(LAST){pumpable};
\item \label{it:lm:blind} $f$ can be computed by a \kl{last $(k{-}1)$-pebble transducer}.
\end{enumerate}
Furthermore, this property is decidable and the construction
is effective.
\end{theorem}

\begin{proof} \Cref{it:lm:blind} $\implies$ \cref{it:lm:growth} is obvious.
\Cref{it:lm:growth} $\implies$ \cref{it:lm:pumpable} is \cref{lem:growth-last}.
Furthermore, \kl(LAST){pumpability} can be tested by an enumeration of $\mu(A^*)$ and $A$.
It remains to show \cref{it:lm:pumpable} $\implies$ \cref{it:lm:blind}
(in an effective fashion): this is the purpose of \cref{ssec:minimize-last}.
\end{proof}

\subsection{Algorithm for removing a recursion layer}

\label{ssec:minimize-last}

\AP
Let $k \ge 2$ and $\utrans$ be a \kl{last $k$-pebble transducer} that is not
\kl(LAST){pumpable}, and that computes $f : A^* \fonc B^*$.
We build a \kl{last $(k{-}1)$-pebble transducer} $\ov{\utrans}$ for $f$.
Let $\mu : (A \uplus \marq{A})^* \fonc \Tb$ be the \kl(LAST){transition morphism}
of $\utrans$. As before (using a \kl{lookaround}), the \kl(LAST){submachines}
of $\ov{\utrans}$ have access to $\forK{\mu}(u)$ on input $u \in A^*$.

\AP
Now, we describe the \kl(LAST){submachines} of $\ov{\utrans}$.
It has  \kl(LAST){submachines} $\reintro*\normalongK{\trans}{\rho}$
for $\trans$ a  \kl(LAST){submachine}  of $\utrans$ and $\rho$
a run of $\trans$, which are described in  \cref{algo:blind-minimization-original}.
Intuitively, these machines mimics the behavior of $\trans$
along the run $\rho$ (which is not necessarily accepting)
of $\trans$ over $\lmark v \rmark$ with $v \in (A \uplus \marq{A})^*$.

\AP
Since they are indexed by a run $\rho$, it may seem that
we create an infinite number of  \kl(LAST){submachines},
but it will not be the case. Indeed, a run $\rho$ will be represented by
its first configuration $(q_1,i_1)$ and last configuration
$(q_n,i_n)$. This information
is sufficient to simulate exactly the two-way moves of $\rho$,
but there is still an unbounded information: the positions
$i_1$ and $i_n$. In fact, the input will be of the form
$v =  \omarK{u}{i}$ and we shall guarantee that the
$i_1$ and $i_n$ can be detected by the \kl{lookaround} if $i$ is marked.
Hence the run $\rho$ will be represented in a bounded
way, independently from the input $v$,
and so that its first and last configurations can be detected
by the \kl{lookaround} of the \kl(LAST){submachine}.

\AP
It follows from \cref{algo:last-minimization-original}
that if $\trans$ is a \kl(LAST){submachine}
of $\utrans$, then for all $v \in (A \cup \marq{A})^*$
and $\rho$ run of $\trans$ on $\lmark v \rmark$,
$\normalongK{\trans}{\rho}~(v)$ is the concatenation
of the outputs produced by (the recursive
calls of) $\trans$ along $\rho$.

\AP
We also define a \kl(LAST){submachine}
$\reintro*\normalongpebK{\trans}{\rho}{i}$
that is similar to $\normalongK{\trans}{\rho}$,
except that it ignores the mark of its input and acts as if it was
in position $i$ (as above for $\rho$, $i$ will be encoded by a bounded
information).

\input{parts/algo-last-minimization-normal.tex}

$\ov{\utrans}$ also has \kl(LAST){submachines} $\reintro*\simLastK{\trans}{\rho}$
for $\trans$ a  \kl(LAST){submachine}  of $\utrans$,
which are described in  \cref{algo:last-minimization-simplify}.
Intuitively, $\simLastK{\trans}{\rho}$ simulates $\trans$ along $\rho$
while trying to inline some recursive calls.
Whenever it is in position $i$ and needs to call recursively
$\ltrans'$ whose \kl(LAST){head}
is $\trans'$, it first \kl{slices} the accepting run $\rho'$ of $\trans'$
on $\lmark \omarK{u}{i} \rmark$, with respect to
$\forK{\mu}{}(u)$ and $i$, as explained
in \cref{def:slice-run} and depicted in \cref{fig:slicing}.
Intuitively, this operation splits $\rho'$ into a bounded number of
runs whose positions either all \kl(POS){observe} $i$,
or $i$ \kl(POS){observes} all of them, or none of these cases
occur (the positions are either $0$, $|u|{+}1$ or \kl(POS){independent} of $i$).

\begin{definition}[Slicing]
\label{def:slice-run}
\AP
Let $u \in A^*$, $\forest \in \FacK{\mu}{}(u)$ and $1 \le i \le |u|$.
We let  $\intro*\ObUpK{i}$ (resp. $\intro*\ObDoK{i}$) be the set of positions
that $i$ \kl(POS){observes} (resp. that  \kl(POS){observe} $i$).\\
Let $\rho = (q_1, i_1) \runs{} \cdots \runs{} (q_n,i_n)$ be a run of a \kl{two-way
transducer} $\trans$ on $\lmark \omarK{u}{i} \rmark$.
We build by induction a sequence $\ell_1, \dots, \ell_{N+1}$ with
$\ell_1 \defined 1$ and:
\begin{itemize}
\item if $\ell_j =  n {+} 1$ then $j \defined N$ and the process ends;
\item else if $i_{\ell_{j}} \in \ObUpK{i}$
(resp. $i_{\ell_{j}} \in \ObDoK{i} \smallsetminus \ObUpK{i}$,
resp. $i_{\ell_{j}} \in [0{:}|u|{+}1] \smallsetminus (\ObUpK{i} \cup \ObDoK{i})$),
then $\ell_{j+1}$ is the largest index such that for all $\ell_{j} \le \ell \le \ell_{j+1}{-}1$,
$i_{\ell} \in \ObUpK{i}$ (resp. $i_{\ell} \in \ObDoK{i} \smallsetminus \ObUpK{i}$,
resp. $i_{\ell} \in [0{:}|u|{+}1] \smallsetminus (\ObUpK{i} \cup \ObDoK{i})$).
\end{itemize}
Finally the \intro{slicing} of $\rho$ ,with respect to $\forest$ and $i$,
is the sequence of runs $\rho_1, \dots, \rho_{N}$ where
$\rho_j \defined (q_{\ell_j}, i_{\ell_j}) \runs{} (q_{\ell_j+1}, i_{\ell_j+1})
\runs{} \cdots \runs{} (q_{\ell_{j+1}-1}, i_{\ell_{j+1}-1})$.
\end{definition}

\vspace*{-0.8cm}

\input{parts/fig-slicing.tex}

{\setlength{\parindent}{0pt}
Now, let $\rho'_1, \dots, \rho'_N$ be  \kl{slicing} of the run $\rho'$ of $\trans'$
on the input $\omarK{u}{i}$. For all $1 \le j \le N$, there are mainly two cases.
Either  the positions of $\rho'_j$ all are in $\ObUpK{i}$ or
$\ObDoK{i}$. In this case, $\simLastK{\trans}{\rho}$  directly inlines $\normalongK{\trans'}{\rho'_j}$
within its own run (i.e. without making a recursive call).
Otherwise, it makes a recursive call to $\simLastK{\trans'}{\rho'_j}$,
except if $\ltrans'$ is a leaf of $\utrans$ (thus $\ltrans' = \trans'$).}

\input{parts/algo-last-minimization-simplify.tex}

Finally, $\ov{\utrans}$  is described as follows: on input 
$u \in A^*$, its \kl(LAST){head} is the \kl(LAST){submachine} $\simLastK{\trans}{\rho}~(u)$,
where $\trans$ is the \kl(LAST){head} of $\utrans$
and $\rho$ is the accepting run of $\trans$ on $\lmark u \rmark$
(represented by the bounded information that it is both initial and final).
As before, we remove the \kl(LAST){submachines}
which are never called in $\ov{\utrans}$.
Observe that we have created a machine with recursion height $k{-}1$\label{sta:last:k}
(because \cref{line:testtest-no2DT}
in \cref{algo:last-minimization-simplify} prevents from calling
a $k$-th layer).

\AP
Let us justify that each $\simLastK{\trans}{\rho}$
can indeed be implemented by a \kl{two-way transducer}.
First, let us observe that since $\forest$ has bounded
height, the number $N$ of \kl{slices} given in \cref{line:N-slice}
of \cref{algo:last-minimization-simplify}
is bounded. Furthermore, we claim that the first and
last positions of each $\rho'_j$ belong to a given
set of bounded size, which can be detected by a \kl{lookaround} which has
access to $i$. For the $\rho'_j$ whose positions are
in $\ObUpK{i}$, this is clear since $|\ObUpK{i}|$ is bounded
(because the frontier of any node is bounded).
For $\ObDoK{i} \smallsetminus \ObUpK{i}$ we use
\cref{lem:struct-O2}, which implies that this set is a bounded
union of intervals. The last case is very similar.

\begin{lemma} \label{lem:struct-O2} Let $1 \le i \le |u|$,
$\nod \defined \originK{\forest}(i)$ and $\nod_{1}$ (resp. $\nod_2$)
be its immediate left (resp. right) sibling (they exist whenever $\nod \in \IteK{\forest}$,
i.e. here $\nod \neq \forest$). Then:
\begin{equalign*}
\ObDoK{i} \smallsetminus \ObUpK{i}  = [\min(\FroK{\forest}(\nod_1)) : \max(\FroK{\forest}(\nod_2))] \smallsetminus \{\FroK{\forest}(\nod_1), \FroK{\forest}(\nod),
\FroK{\forest}(\nod_2)\}.
\end{equalign*}
\end{lemma}
This analysis justifies why each $\rho'_j$ can be encoded in a bounded way.
Now, we show how to implement the inlinings while
using $i$ as the current position:
\begin{itemize}
\item if $i_1, \dots, i_n \in \ObUpK{i}$,
then $n$ is bounded (because $|\ObUpK{i}|$ is bounded).
We can thus inline $\normalongK{\trans'}{{\rho'_j}}~(\omarK{u}{i})$
while staying in position $i$. However, when $\trans'$ calls
some $\ltrans''$ (of \kl(LAST){head} $\trans''$) on
position $i_{\ell}$, we would need to call
$\normalongK{\trans''}{\rho''}(\omarK{u}{i_{\ell}})$
(where $\rho''$ is the accepting run of $\trans''$ along
$\lmark \omarK{u}{i_{\ell}} \rmark$). But we cannot do this operation,
since we are in position $i$ and not in $i_{\ell}$. The solution is that the inlined code
calls $\normalongpebK{\trans''}{\rho''}{i_{\ell}}(\omarK{u}{i})$
instead, which simulates an accepting run $\rho''$
of $\trans$ on $\omarK{u}{i_{\ell}}$, even if its input is $\omarK{u}{i}$.
Note that $i_{\ell}$ can be represented as a bounded information
and recovered by a \kl{lookaround} given $\omarK{u}{i}$ as input,
since $i$ \kl(POS){observes} $i_{\ell}$;

 \item if $i_1, \dots, i_n \in \ObDoK{i} \smallsetminus \ObUpK{i}$, then
 the nodes $\originK{\forest}(i_{1}), \dots, \originK{\forest}(i_{n})$
 are \linebreak roughly below $\originK{\forest}(i)$ in $\forest$ (see \cref{fig:observation}).
 We inline $\normalongK{\trans'}{{\rho'_j}}~(\omarK{u}{i})$,
 by moving along $i_1, \dots, i_n$ as $\rho'_j$ does.
We can keep track of the height of $\originK{\forest}(i)$
 above the current $\originK{\forest}(i_{\ell})$
 (it is a bounded information). With the
\kl{lookaround}, we can detect the end of $\rho'_j$, and
 go back to position $i$.

\end{itemize}

\AP
It remains to justify that $\ov{\utrans}$
is correct. For this, we only need to show that
when it reaches \cref{line:key-blind-minimization}
in \cref{algo:last-minimization-simplify}, the output of
$\trans'$ along $\rho'_j$ is indeed empty.
Otherwise, the conditions of \cref{lem:k-nodes-last-key}
would hold \label{sta:last:2}(since we never execute two successive recursive
calls in \kl(POS){dependent} positions). 
It provides a contradiction.

\begin{lemma}[Key lemma]%
\label{lem:k-nodes-last-key}
Let $u \in A^*$ and $\forest \in \FacK{\mu}{}(u)$.
Assume that there exists a sequence $\trans_1, \dots, \trans_k$
of \kl(LAST){submachines}
of $\utrans$ and a sequence of positions 
$1 \le i_1, \dots, i_k \le |u|$ such that:
\begin{itemize}
\item $\trans_1$ is the \kl(LAST){head} of $\utrans$;
\item $|\valK{\trans_1}{u}(i_1)|_{\trans_2} \neq 0$ and
$\valK{\trans_k}{\omarK{u}{i_{k-1}}}(i_k) \neq \movi$;
\item for all $2 \le j \le k{-}1$, $| \valK{\trans_j}{\omarK{u}{i_{j-1}}}(i_j) |_{\trans_{j+1}} \neq 0$;
\item for all $1 \le j \le k{-}1$, $\originK{\forest}(i_j)$
and $\originK{\forest}(i_{j+1})$ are \kl(NOD){independent};
\end{itemize}
Then $\utrans$ is \kl(LAST){pumpable}.
\end{lemma}

\begin{proof}[idea] As for \cref{lem:k-nodes-blind-key},
the key observation is that \kl(LAST){pumpability}
follows as soon as the nodes $\originK{\forest}(i_j)$ are
pairwise \kl(NOD){independent}. Furthermore, this condition
can be obtained by duplicating some nodes  in $\forest$.
\end{proof}

\section{Making the two last pebbles visible}

\label{sec:last-last}

\AP
We can define a similar model to that of \kl{last $k$-pebble
transducer}, which sees the two last calling positions instead of only the previous one.
Let us name this model a \intro{last-last $k$-pebble transducer}.
A very natural question is to know whether we can
show an analog of \cref{the:minimize-main} for these machines.

Note that for $k=1,2$ and $3$, a \kl{last-last $k$-pebble transducer} is
exactly the same as a \kl{$k$-pebble  transducer}. Hence the function
$\isqK$ of \cpageref{page:isq} is such that $|\isqK(u)| = \bigO(|u|^2)$
and can be computed by  a \kl{last-last $3$-pebble transducer}, but it cannot be computed by
a \kl{last-last $2$-pebble transducer}. It follows
that the connection between minimal recursion height and growth of the output
fails. However, this result is somehow artificial.
Indeed, a \kl{last-last $2$-pebble transducer} is 
a degenerate case, since it can only see one last pebble.
More interestingly, we show that the connection fails
for arbitrary heights.

\begin{theorem}
\label{the:fail-last-last}
For all $k \ge 2$, there exists a function $f:A^* \fonc B^*$ such that 
$|f(u)| = \bigO(|u|^2)$ and that can be computed by
a \kl{last-last $(2k{+}1)$-pebble transducer},
but not by a \kl{last-last $2k$-pebble transducer}.
\end{theorem}

\begin{proof}[idea] We re-use a counterexample introduced by
Boja{\'{n}}czyk in \cite{bojanczyk2022growth} to show a similar
failure result for the model of \kl{$k$-pebble transducers}.
\end{proof}

\section{Outlook}

This paper somehow settles the discussion concerning the variants
of \kl{pebble transducers} for which the minimal recursion height
only depends on the growth of the output. As soon as two marks are visible,
the combinatorics of the output also has to be taken into account,
hence minimizing the recursion height in this case (e.g. for \kl{last-last pebble transducers})
 seems hard with the current tools.

As observed in \cite{engelfriet2007xml}, one can extend
\kl{last pebble transducers} by allowing the recursion height
to be unbounded  (in the spirit of
\intro{marble transducers} \cite{doueneau2020register}).
This model enables to produce outputs whose size grows exponentially
in the size of the input. A natural question is to know
whether a function computed by this model,
but whose output size is polynomial, can in fact be computed with
a recursion stack of bounded height (i.e. by a \kl{last $k$-pebble transducer}).

\subsubsection{Acknowledgements.} The author is grateful to Tito Nguy{\^{e}}n
for suggesting the study of the recursion height for \kl{last pebble transducers}.

%
%
%

\newpage

 \bibliographystyle{splncs04}
\bibliography{globals/last-pebble}

\appendix

\input{parts/appendix.tex}

\end{document}

%% file: globals/imports.tex
\usepackage{amsmath,mathrsfs}
\usepackage[UKenglish]{babel}
\usepackage{graphicx}
\usepackage{xspace}
\usepackage{color} 
\usepackage{subcaption} 
\usepackage{xy}
\usepackage{tikz-cd}

\definecolor{myBlue}{HTML}{88C0D0}
\definecolor{myDarkBlue}{HTML}{5E81AC}
\definecolor{myTeal}{HTML}{8FBCBB}
\definecolor{myOrange}{HTML}{D08770}
\definecolor{myGreen}{HTML}{A3BE8C}
\definecolor{myRed}{HTML}{BF616A}
\definecolor{myYellow}{HTML}{EBCB8B}
\definecolor{myPurple}{HTML}{B48EAD}
\definecolor{myBlack}{HTML}{3B4252}
\colorlet{myDeepPurple}{myPurple!150}
\colorlet{myDeepBlue}{myDarkBlue!150}
\colorlet{myGray}{myBlack!80}
\colorlet{myLightGray}{myBlack!30}

\usepackage{mathtools}
\usepackage{mathrsfs}
\usepackage{alltt}
\usepackage{epigraph}
\usepackage{tikz}
\usetikzlibrary{automata,positioning,arrows}
\usepackage{amsfonts, amssymb,dsfont, mathrsfs}
\usepackage{mathtools}

\usepackage{hyperref}
\usepackage[ruled,linesnumbered,commentsnumbered]{algorithm2e}
\usepackage{stmaryrd}
\usepackage{pifont,enumitem}
\usepackage{soulutf8}
\usepackage{yhmath}
\usepackage{bm}

\usepackage[cleveref,hyperref, xcolor, notion,makeidx,paper]{knowledge}
\input{globals/knowledges.kl}
\knowledgestyle{intro notion}{color=myDeepBlue,boldface }
\knowledgestyle{notion}{color=myGray!120}
\makeindex

\newenvironment{equalign*}{%
\begin{equation*}
\begin{aligned}%
}{%
\end{aligned}
\end{equation*}}

\setlist{nosep}

%% file: globals/macros.tex
\newcommand{\teq}[1]{\text{\small #1}}

\newcommand{\mb}[1]{\mathbb{#1}}
\newcommand{\mc}[1]{\mathcal{#1}}
\newcommand{\mf}[1]{\mathfrak{#1}}

\newcommand{\inter}[2]{[{#1}{:}{#2}]}

\newcommand{\movi}{\varepsilon}
\newcommand{\vide}{\varnothing}
\renewcommand{\implies}{\Rightarrow}
\newcommand{\defined}{\coloneqq}
\renewcommand{\phi}{\varphi}
\renewcommand{\le}{\leqslant}
\renewcommand{\ge}{\geqslant}

\renewcommand{\epsilon}{\operatorname{\normalfont \textcolor{green}{%
\textsf{FORBIDDEN}}}}
\newcommand{\Nat}{\mb{N}}

\newcommand{\Mono}{\mb{M}}

\newcommand{\Tb}{\mb{T}}

\newcommand{\parfonc}{\rightharpoonup}
\newcommand{\fonc}{\rightarrow}

\newcommand{\ov}[1]{\overline{#1}}

\newcommand{\lefttree}{\langle}
\newcommand{\righttree}{\rangle}
\newcommand{\tree}[1]{\lefttree #1 \righttree}

\newcommand{\transfont}[1]{\mathscr{#1}}
\newcommand{\btrans}{\transfont{B}}

\newcommand{\ltrans}{\transfont{L}}
\newcommand{\trans}{\transfont{T}}

\newcommand{\utrans}{\transfont{U}}

\newcommand{\Mcont}{\mc{M}}
\newcommand{\Ccont}{\mc{C}}
\newcommand{\Mblock}[2]{\Mcont_{#1}^{#2}}

\newcommand{\lmove}{\triangleleft}
\newcommand{\rmove}{\triangleright}
\newcommand{{\lmark}}{{\vdash}}
\newcommand{\rmark}{{\dashv}}

\newcommand{\runs}[1]{\mathchoice{\xrightarrow{#1}}{\xrightarrow{\smash{\lower1pt\hbox{$\scriptstyle #1$}}}}{\xrightarrow{#1}}{\xrightarrow{#1}}}

\newcommand{\bigO}{\mc{O}}
\newcommand{\bigT}{\Theta}

\newcommand{\marq}[1]{\underline{#1}}

\newcommand{\forest}{\mc{F}}

\newcommand{\apar}[1]{{\widehat{#1}}}
\newcommand{\nod}{\mf{t}}

\newcommand{\wt}[1]{\widetilde{#1}}

\newcommand{\tred}[1]{\textcolor{myRed}{#1}}

\newcommand{\cro}[1]{ \tred{\bm{\llbracket} #1\bm{\rrbracket}}}

\newcommand{\Ope}[1]{{\normalfont{\textsf{#1}}}}

\knowledgenewrobustcmd\ulsq{\cmdkl{\operatorname{\normalfont\mathsf{unmarked-square}}}}
\knowledgenewrobustcmd\omarK[2]{#1\cmdkl{\bullet} #2}
\knowledgenewrobustcmd\sqK{\cmdkl{\operatorname{\normalfont\mathsf{square}}}}
\knowledgenewrobustcmd\parikh{\cmdkl{\operatorname{\normalfont\mathsf{parikh}}}}
\knowledgenewrobustcmd\itpow[1]{\cmdkl{\operatorname{\normalfont\textsf{powers}}}^{#1}}
\knowledgenewrobustcmd\FroK[1]{\cmdkl{\operatorname{\normalfont \textsf{Fr}}}^{#1}}
\knowledgenewrobustcmd\SkeK[1]{\cmdkl{\operatorname{\normalfont\textsf{Skel}}}^{#1}}
\knowledgenewrobustcmd\originK[1]{\cmdkl{\operatorname{\normalfont\textsf{origin}}}^{#1}}
\knowledgenewrobustcmd\linK[1]{\cmdkl{\operatorname{\normalfont\textsf{lin}}}^{#1}}
\knowledgenewrobustcmd\IteK[1]{\cmdkl{\operatorname{\normalfont \textsf{Iter}}}^{#1}}
\knowledgenewrobustcmd\NodK[1]{\cmdkl{\operatorname{\normalfont \textsf{Nodes}}}^{#1}}
\knowledgenewrobustcmd\FacK[2]{\cmdkl{\operatorname{\normalfont \textsf{Forests}}}_{#1}^{#2}}
\knowledgenewrobustcmd\isqK{\cmdkl{\operatorname{\normalfont  \textsf{inner-squaring}}}}
\knowledgenewrobustcmd\mapRevK{\cmdkl{\operatorname{\normalfont  \textsf{map-reverse}}}}
\knowledgenewrobustcmd\forK[1]{\cmdkl{\operatorname{\normalfont  \textsf{forest}}}_{#1}}
\knowledgenewrobustcmd\ObUpK[1]{\cmdkl{\uparrow #1}}
\knowledgenewrobustcmd\ObDoK[1]{\cmdkl{\downarrow #1}}

\knowledgenewrobustcmd\crossK[2]{\cmdkl{\Ope{\textsf{cross}}}_{#1}^{#2}}
\knowledgenewrobustcmd\valK[2]{\cmdkl{\Ope{\textsf{prod}}}_{#1}^{#2}}
\knowledgenewrobustcmd\zebraK[1]{\cmdkl{\Ope{\textsf{alternating-square}}}^{#1}}

\knowledgenewrobustcmd\simBlindK[1]{%
\cmdkl{{\Ope{accelerate-}#1}}}
\knowledgenewrobustcmd\simLastK[2]{%
\cmdkl{{\Ope{accelerate-}#1\Ope{-along-}#2}}}

\knowledgenewrobustcmd\normalK[1]{%
\cmdkl{{\Ope{old-}#1}}}
\knowledgenewrobustcmd\normalongK[2]{%
\cmdkl{{\Ope{old-}#1%
\Ope{-along-}#2}}}
\knowledgenewrobustcmd\normalongpebK[3]{%
\cmdkl{{\Ope{normal-}#1%
\Ope{-along-}#2\Ope{-pebble-}#3}}}

\knowledgenewrobustcmd\outK[2]{%
\cmdkl{{\Ope{output-of-}#1\Ope{-in-}#2}}}

%% file: parts/fig-pebble-transducer.tex
\begin{figure}[h!]

\centering
\scalebox{0.8}{
\begin{tikzpicture}{scale=1}

	\newcommand{\coulun}{myDeepPurple}
	\newcommand{\coulde}{myDarkBlue}
	\newcommand{\coultr}{myRed}
	\newcommand{\texte}{\small \bfseries \sffamily \mathversion{bold} }

	\draw (-0.25,4) rectangle (8.25,4.5);
	\node[above] at (6.2,4) {$\text{\small Input word}$};
	\node[above] at (0.05,4) {$\lmark$};
	\node[above] at (7.95,4) {$\rmark$};

	\node[above] at (-1.9,3.25) {\textcolor{\coulun}{\text{ \bfseries Main machine}}};
	\draw[-,line width = 2pt,\coulun](0,3.8) -- (7,3.8);
	\draw[-,line width = 2pt, \coulun] (7,3.8) arc (90:-90:0.2);
	\draw[-,line width = 2pt,\coulun](7,3.4) -- (1.5,3.4);	
	\draw[-, line width = 2pt,\coulun] (1.5,3.4) arc (90:270:0.2);
	\draw[-,line width = 2pt,\coulun](1.5,3) -- (2.5,3);	
	\fill[fill = \coulun,even odd rule] (2.5,3) circle (0.08);


	\draw (-0.25,2.5) rectangle (8.25,2);
	\node[above] at (6.2,2) {\text{\small Input word}};
	\node[above] at (0.05,2) {$\lmark$};
	\node[above] at (7.95,2) {$\rmark$};
	\fill[fill = \coulun] (2.5,2.25) circle (0.15);

	\node[above] at (-2.5,1.35) {\textcolor{\coulde}{\text{\bfseries%
	Submachine called in $\textcolor{\coulun}{\bullet}$}}};
	\draw[-,line width = 2pt,\coulde](0,1.8) -- (3,1.8);
	\draw[-,line width = 2pt, \coulde] (3,1.8) arc (90:-90:0.2);
	\draw[-,line width = 2pt,\coulde](3,1.4) -- (1,1.4);	
	\draw[-,line width = 2pt,\coulde] (1,1.4) arc (90:270:0.2);
	\draw[-,line width = 2pt,\coulde](1,1) -- (5,1);	
	\fill[fill = \coulde,even odd rule] (5,1) circle (0.08);

        \draw[->,line width = 2pt,\coulun,dashed](2.5,3) to[out= -120, in = 60] (0,1.9);	
	
	\node[above,\coulun] at (2.5,1.7) {$\substack{\text{pebble}}$};


	\draw (-0.25,0.5) rectangle (8.25,0);
	\node[above] at (6.2,0) {\text{\small Input word}};
	\node[above] at (0.05,0) {$\lmark$};
	\node[above] at (7.95,0) {$\rmark$};
	\fill[fill = \coulde] (5,0.25) circle (0.15);
	\fill[fill = \coulun] (2.5,0.25) circle (0.15);

	\node[above] at (-2.5,-0.75) {\textcolor{\coultr}
	{\text{ \bfseries Submachine called in $\textcolor{\coulde}{\bullet}$} }};
	\draw[-,line width = 2pt,\coultr](0,-0.2) -- (1,-0.2);
	\draw[-,line width = 2pt, \coultr](1,-0.2) arc (90:-90:0.2);
	\draw[-,line width = 2pt,\coultr] (1,-0.6) -- (0.5,-0.6);	
	\draw[-, line width = 2pt,\coultr] (0.5,-0.6) arc (90:270:0.2);
	\draw[-,line width = 2pt,\coultr] (0.5,-1) -- (6,-1);	
	\fill[fill = \coultr,even odd rule] (6,-1) circle (0.08);

        \draw[->,line width = 2pt,\coulde,dashed](5,1) to[out= -150, in = 40] (0,-0.1);	
	
	\node[above,\coulde] at (5,-0.3) {$\substack{\text{pebble}}$};
	\node[above,\coulun] at (2.5,-0.3) {$\substack{\text{pebble}}$};

\end{tikzpicture}}

\caption{\label{fig:pebble} Behavior of a \kl{$3$-pebble transducer}.}

\end{figure}

%% file: parts/fig-blind-transducer.tex
\begin{figure}[h!]

\centering
\scalebox{0.8}{
\begin{tikzpicture}{scale=1}

	\newcommand{\coulun}{myDeepPurple}
	\newcommand{\coulde}{myDarkBlue}
	\newcommand{\coultr}{myRed}
	\newcommand{\texte}{\small \bfseries \sffamily \mathversion{bold} }

	\draw (-0.25,4) rectangle (8.25,4.5);
	\node[above] at (6.2,4) {$\text{\small Input word}$};
	\node[above] at (0.05,4) {$\lmark$};
	\node[above] at (7.95,4) {$\rmark$};

	\node[above] at (-1.9,3.25) {\textcolor{\coulun}{\text{ \bfseries Main machine}}};
	\draw[-,line width = 2pt,\coulun](0,3.8) -- (7,3.8);
	\draw[-,line width = 2pt, \coulun] (7,3.8) arc (90:-90:0.2);
	\draw[-,line width = 2pt,\coulun](7,3.4) -- (1.5,3.4);	
	\draw[-, line width = 2pt,\coulun] (1.5,3.4) arc (90:270:0.2);
	\draw[-,line width = 2pt,\coulun](1.5,3) -- (2.5,3);	
	\fill[fill = \coulun,even odd rule] (2.5,3) circle (0.08);


	\draw (-0.25,2.5) rectangle (8.25,2);
	\node[above] at (6.2,2) {\text{\small Input word}};
	\node[above] at (0.05,2) {$\lmark$};
	\node[above] at (7.95,2) {$\rmark$};

	\node[above] at (-2.5,1.35) {\textcolor{\coulde}{\text{\bfseries%
	Submachine called in $\textcolor{\coulun}{\bullet}$}}};
	\draw[-,line width = 2pt,\coulde](0,1.8) -- (3,1.8);
	\draw[-,line width = 2pt, \coulde] (3,1.8) arc (90:-90:0.2);
	\draw[-,line width = 2pt,\coulde](3,1.4) -- (1,1.4);	
	\draw[-,line width = 2pt,\coulde] (1,1.4) arc (90:270:0.2);
	\draw[-,line width = 2pt,\coulde](1,1) -- (5,1);	
        \draw[->,line width = 2pt,\coulun,dashed](2.5,3) to[out= -120, in = 60] (0,1.9);	
	\fill[fill = \coulde,even odd rule] (5,1) circle (0.08);


	\draw (-0.25,0.5) rectangle (8.25,0);
	\node[above] at (6.2,0) {\text{\small Input word}};
	\node[above] at (0.05,0) {$\lmark$};
	\node[above] at (7.95,0) {$\rmark$};

	\node[above] at (-2.5,-0.75) {\textcolor{\coultr}
	{\text{ \bfseries Submachine called in $\textcolor{\coulde}{\bullet}$} }};
	\draw[-,line width = 2pt,\coultr](0,-0.2) -- (1,-0.2);
	\draw[-,line width = 2pt, \coultr](1,-0.2) arc (90:-90:0.2);
	\draw[-,line width = 2pt,\coultr] (1,-0.6) -- (0.5,-0.6);	
	\draw[-, line width = 2pt,\coultr] (0.5,-0.6) arc (90:270:0.2);
	\draw[-,line width = 2pt,\coultr] (0.5,-1) -- (6,-1);	
	\fill[fill = \coultr,even odd rule] (6,-1) circle (0.08);

        \draw[->,line width = 2pt,\coulde,dashed](5,1) to[out= -150, in = 40] (0,-0.1);

\end{tikzpicture}}

\caption{\label{fig:3-blind-transducer} Behavior of a \kl{blind $3$-pebble transducer}.}

\end{figure}

%% file: parts/fig-last-transducer.tex
\begin{figure}[h!]

\centering
\scalebox{0.8}{
\begin{tikzpicture}{scale=1}

	\newcommand{\coulun}{myDeepPurple}
	\newcommand{\coulde}{myDarkBlue}
	\newcommand{\coultr}{myRed}
	\newcommand{\texte}{\small \bfseries \sffamily \mathversion{bold} }

	\draw (-0.25,4) rectangle (8.25,4.5);
	\node[above] at (6.2,4) {$\text{\small Input word}$};
	\node[above] at (0.05,4) {$\lmark$};
	\node[above] at (7.95,4) {$\rmark$};

	\node[above] at (-1.9,3.25) {\textcolor{\coulun}{\text{ \bfseries Main machine}}};
	\draw[-,line width = 2pt,\coulun](0,3.8) -- (7,3.8);
	\draw[-,line width = 2pt, \coulun] (7,3.8) arc (90:-90:0.2);
	\draw[-,line width = 2pt,\coulun](7,3.4) -- (1.5,3.4);	
	\draw[-, line width = 2pt,\coulun] (1.5,3.4) arc (90:270:0.2);
	\draw[-,line width = 2pt,\coulun](1.5,3) -- (2.5,3);	
	\fill[fill = \coulun,even odd rule] (2.5,3) circle (0.08);


	\draw (-0.25,2.5) rectangle (8.25,2);
	\node[above] at (6.2,2) {\text{\small Input word}};
	\node[above] at (0.05,2) {$\lmark$};
	\node[above] at (7.95,2) {$\rmark$};
	\fill[fill = \coulun] (2.5,2.25) circle (0.15);

	\node[above] at (-2.5,1.35) {\textcolor{\coulde}{\text{\bfseries%
	Submachine called in $\textcolor{\coulun}{\bullet}$}}};
	\draw[-,line width = 2pt,\coulde](0,1.8) -- (3,1.8);
	\draw[-,line width = 2pt, \coulde] (3,1.8) arc (90:-90:0.2);
	\draw[-,line width = 2pt,\coulde](3,1.4) -- (1,1.4);	
	\draw[-,line width = 2pt,\coulde] (1,1.4) arc (90:270:0.2);
	\draw[-,line width = 2pt,\coulde](1,1) -- (5,1);	
	\fill[fill = \coulde,even odd rule] (5,1) circle (0.08);

        \draw[->,line width = 2pt,\coulun,dashed](2.5,3) to[out= -120, in = 60] (0,1.9);	
	
	\node[above,\coulun] at (2.5,1.7) {$\substack{\text{pebble}}$};


	\draw (-0.25,0.5) rectangle (8.25,0);
	\node[above] at (6.2,0) {\text{\small Input word}};
	\node[above] at (0.05,0) {$\lmark$};
	\node[above] at (7.95,0) {$\rmark$};
	\fill[fill = \coulde] (5,0.25) circle (0.15);

	\node[above] at (-2.5,-0.75) {\textcolor{\coultr}
	{\text{ \bfseries Submachine called in $\textcolor{\coulde}{\bullet}$} }};
	\draw[-,line width = 2pt,\coultr](0,-0.2) -- (1,-0.2);
	\draw[-,line width = 2pt, \coultr](1,-0.2) arc (90:-90:0.2);
	\draw[-,line width = 2pt,\coultr] (1,-0.6) -- (0.5,-0.6);	
	\draw[-, line width = 2pt,\coultr] (0.5,-0.6) arc (90:270:0.2);
	\draw[-,line width = 2pt,\coultr] (0.5,-1) -- (6,-1);	
	\fill[fill = \coultr,even odd rule] (6,-1) circle (0.08);

        \draw[->,line width = 2pt,\coulde,dashed](5,1) to[out= -150, in = 40] (0,-0.1);	
	
	\node[above,\coulde] at (5,-0.3) {$\substack{\text{pebble}}$};

\end{tikzpicture}}

\caption{\label{fig:3-last-transducer} Behavior of a \kl{last $3$-pebble transducer}.}

\end{figure}

%% file: parts/fig-factorization-forest.tex
\begin{figure}[h!]
	\newcommand{\couleur}{myDarkBlue}
	\centering
	\scalebox{0.8}{
	\begin{tikzpicture}

		\newcommand{\texte}{\small \bfseries \sffamily \mathversion{bold} }

		\node[above] at  (0,0)  {$-1$};
		\node[above] at  (1,0)  {$-1$};
		\node[above] at  (2,0)  {$0$};
		\node[above] at  (3,0)  {$-1$};
		\node[above] at  (4,0)  {$0$};
		\node[above] at  (5,0)  {$0$};
		\node[above] at  (6,0)  {$0$};
		\node[above] at  (7,0)  {$0$};
		\node[above] at  (8,0)  {$0$};
		\node[above] at  (9,0)  {$0$};

		\draw[very thick, double] (4,0.975) -- (8,0.975);
		\draw[very thick,] (3,1.5) -- (6,1.5);
		\draw[very thick,double] (2,1.975) -- (9,1.975);
		\draw[very thick,] (0,2) -- (1,2);
		\draw[very thick,] (0.5,2.5) -- (6.5,2.5);

		\draw[very thick,] (0,2) -- (0,0.5);
		\draw[very thick,] (1,2) -- (1,0.5);
		\draw[very thick,] (0.5,2) -- (0.5,2.5);

		\draw[very thick,] (2,2) -- (2,0.5);
		\draw[very thick,] (9,2) -- (9,0.5);

		\draw[very thick,] (6.5,2) -- (6.5,2.5);
		\draw[very thick,] (3.5,2.5) -- (3.5,2.75);

		\draw[very thick,] (3,1.5) -- (3,0.5);
		\draw[very thick,] (4,1) -- (4,0.5);
		\draw[very thick,] (5,1) -- (5,0.5);
		\draw[very thick,] (6,1.5) -- (6,0.5);
		\draw[very thick,] (7,1) -- (7,0.5);
		\draw[very thick,] (8,1) -- (8,0.5);

		\draw[very thick,] (4.5,2) -- (4.5,1.5);
		
		\fill[fill = \couleur,even odd rule] (4.5,1.5) circle (0.15);
		\fill[fill = \couleur,even odd rule] (6,1)  circle (0.15);
		\fill[fill = \couleur,even odd rule]  (3,0.5) circle (0.15);
		\fill[fill = \couleur,even odd rule] (4,0.5) circle (0.15);
		\fill[fill = \couleur,even odd rule] (8,0.5) circle (0.15);

	\end{tikzpicture}}

	\caption{\label{fig:ex:skel}  $ \forest \in \FacK{\mu}{}((-1)(-1)0(-1)000000)$
		and a \kl{skeleton}.}
\end{figure}

%% file: parts/fig-observable.tex
\begin{figure}[h!]

\centering
	\newcommand{\coulun}{myDeepPurple}
	\newcommand{\coulde}{myRed}
	\newcommand{\coultr}{myDarkBlue}

\scalebox{0.8}{
\begin{tikzpicture}{scale=1}

	\draw[-,line width = 2pt,\coulun](0,0) -- (-0.5,-2) -- (0.5,-2) -- cycle;
	
	\node[above] at (0,-2.5) {\textcolor{\coulun}{\footnotesize \bfseries Nodes that %
	observe \textcolor{\coultr}{$\bullet$}}};
	
	\node[above] at (3.8,1.65) {\textcolor{\coulde}{\footnotesize \bfseries %
	\textcolor{\coultr}{$\bullet$} observes these nodes}};

	\draw[double, very thick] (-2,0.5) -- (2,0.5);
	\draw[very thick] (-1.2,0.5) -- (-1.2,0);
	\draw[very thick] (1.2,0.5) -- (1.2,0);
	\draw[very thick] (0,0.5) -- (0,0);

	\draw[very thick] (0,0.5) -- (0,1) -- (3,1) -- (3,0.5);
	\draw[very thick] (1.5,1) -- (1.5,2);
	\draw[double, very thick] (-5,1.5) -- (7,1.5);
	\draw[very thick] (-2.75,1.5) -- (-2.75,1);
	\draw[very thick] (4.75,1.5) -- (4.75,1);
	\draw[very thick] (-3.75,1.5) -- (-3.75,1);
	\draw[very thick] (5.75,1.5) -- (5.75,1);
	
	\draw[-,line width = 2pt,\coulun](1.2,0) -- (0.7,-2) -- (1.7,-2) -- cycle;

	\draw[-,line width = 2pt,\coulun](-1.2,0) -- (-0.7,-2) -- (-1.7,-2) -- cycle;

	\fill[fill = \coulde,even odd rule] (0,0) circle (0.2);
	\fill[fill = \coultr,even odd rule] (0,0) circle (0.15);
	\fill[fill = \coulde,even odd rule] (1.2,0) circle (0.15);
	\fill[fill = \coulde,even odd rule] (-1.2,0) circle (0.15);
	
	\fill[fill = \coulde,even odd rule] (0,0.5) circle (0.15);
	\fill[fill = \coulde,even odd rule] (1.5,1) circle (0.15);
	\fill[fill = \coulde,even odd rule] (3,0.5) circle (0.15);
	\fill[fill = \coulde,even odd rule] (1.5,2) circle (0.15);
	\fill[fill = \coulde,even odd rule] (-2.75,1) circle (0.15);
	\fill[fill = \coulde,even odd rule] (4.75,1) circle (0.15);

\end{tikzpicture}}

\caption{\label{fig:observation} Nodes that \kl(NOD){observe}
\textcolor{\coultr}{$\bullet$} and that \textcolor{\coultr}{$\bullet$} \kl(NOD){observes}}

\end{figure}

%% file: parts/fig-blind-pumpable.tex
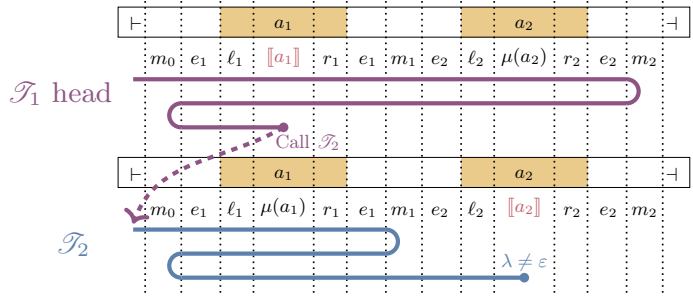
\begin{figure}[h!]

\centering
\scalebox{0.8}{
\begin{tikzpicture}{scale=1}

	\newcommand{\coulun}{myDeepPurple}
	\newcommand{\coulde}{myDarkBlue}
	\newcommand{\coultr}{myYellow}

	\fill[\coultr] (1.45,5) rectangle (3.55,4.5);
	\fill[\coultr] (5.45,5) rectangle (7.55,4.5);
	\draw (-0.25,5) rectangle (9.25,4.5);
	\node[above] at (2.5,4.5) {$a_1$};
	\node[above] at (6.5,4.5) {$a_2$};
	\node[above] at (0.05,4.5) {$\lmark$};
	\node[above] at (8.95,4.5) {$\rmark$};

	\node[above] at (0.5,3.9) {$m_0$};
	\node[above] at (1.1,3.9) {$e_1$};	
	\node[above] at (1.7,3.9) {$\ell_1$};	
	\node[above] at (2.5,3.9) {$\cro{a_1}$};
	\node[above] at (3.3,3.9) {$r_1$};
	\node[above] at (3.9,3.9) {$e_1$};
	\node[above] at (4.5,3.9) {$m_1$};
	\node[above] at (5.1,3.9) {$e_2$};
	\node[above] at (5.7,3.9) {$\ell_2$};
	\node[above] at (6.5,3.9) {$\mu(a_2)$};
	\node[above] at (7.3,3.9) {$r_2$};
	\node[above] at (7.9,3.9) {$e_2$};
	\node[above] at (8.5,3.9) {$m_2$};
	

	\fill[\coultr] (1.45,2.5) rectangle (3.55,2);
	\fill[\coultr] (5.45,2.5) rectangle (7.55,2);
	\draw (-0.25,2.5) rectangle (9.25,2);
	\node[above] at (2.5,2) {$a_1$};
	\node[above] at (6.5,2) {$a_2$};
	\node[above] at (0.05,2) {$\lmark$};
	\node[above] at (8.95,2) {$\rmark$};
	
	\node[above] at (0.5,1.4) {$m_0$};
	\node[above] at (1.1,1.4) {$e_1$};	
	\node[above] at (1.7,1.4) {$\ell_1$};	
	\node[above] at (2.5,1.4) {$\mu(a_1)$};
	\node[above] at (3.3,1.4) {$r_1$};
	\node[above] at (3.9,1.4) {$e_1$};
	\node[above] at (4.5,1.4) {$m_1$};
	\node[above] at (5.1,1.4) {$e_2$};
	\node[above] at (5.7,1.4) {$\ell_2$};
	\node[above] at (6.5,1.4) {$\cro{a_2}$};
	\node[above] at (7.3,1.4) {$r_2$};
	\node[above] at (7.9,1.4) {$e_2$};
	\node[above] at (8.5,1.4) {$m_2$};


	\draw[dotted,  thick] (0.2,5.1) -- (0.2,0.2);
	\draw[dotted,  thick] (0.8,5.1) -- (0.8,0.2);
	\draw[dotted,  thick] (1.45,5.1) -- (1.45,0.2);
	
	\draw[dotted,  thick] (2,5.1) -- (2,0.2);
	\draw[dotted,  thick] (3,5.1) -- (3,0.2);
	\draw[dotted,  thick] (3.55,5.1) -- (3.55,0.2);
	\draw[dotted,  thick] (4.2,5.1) -- (4.2,0.2);

	\draw[dotted,  thick] (7,5.1) -- (7,0.2);
	\draw[dotted,  thick] (6,5.1) -- (6,0.2);
	\draw[dotted,  thick] (5.45,5.1) -- (5.45,0.2);
	\draw[dotted,  thick] (4.8,5.1) -- (4.8,0.2);

	\draw[dotted,  thick] (7.55,5.1) -- (7.55,0.2);
	\draw[dotted,  thick] (8.2,5.1) -- (8.2,0.2);
	\draw[dotted,  thick] (8.8,5.1) -- (8.8,0.2);

	\node[above] at (-1.3,3.25) {\scalebox{1.5}{%
	\textcolor{\coulun}{\text{ {\bfseries $\trans_1$} head}}}};
	\draw[-,line width = 2pt,\coulun](0,3.8) -- (8.2,3.8);
	\draw[-,line width = 2pt, \coulun] (8.2,3.8) arc (90:-90:0.2);
	\draw[-,line width = 2pt,\coulun](8.2,3.4) -- (0.8,3.4);	
	\draw[-, line width = 2pt,\coulun] (0.8,3.4) arc (90:270:0.2);
	\draw[-,line width = 2pt,\coulun](0.8,3) -- (2.5,3);	
	\fill[fill = \coulun,even odd rule] (2.5,3) circle (0.08);

	\node[above] at (-1,0.75) {\scalebox{1.5}{%
	\textcolor{\coulde}{$\trans_2$}}};
	\draw[-,line width = 2pt,\coulde](0,1.3) -- (4.2,1.3);
	\draw[-,line width = 2pt, \coulde] (4.2,1.3) arc (90:-90:0.2);
	\draw[-,line width = 2pt,\coulde](4.2,0.9) -- (0.8,0.9);	
	\draw[-,line width = 2pt,\coulde] (0.8,0.9) arc (90:270:0.2);
	\draw[-,line width = 2pt,\coulde](0.8,0.5) -- (6.5,0.5);	
	\fill[fill = \coulde,even odd rule] (6.5,0.5) circle (0.08);
	\node[above] at (6.5,0.5) {\textcolor{\coulde}{\small $\lambda \neq \movi$}};
	\node[above] at (2.9,2.5) {\textcolor{\coulun}{\small Call $\trans_2$}};

        \draw[->,line width = 2pt,\coulun,dashed](2.5,3) to[out= -150, in = 70] (0,1.4);

\end{tikzpicture}}

\caption{\label{fig:blind-pumpable} \kl(BLIND){Pumpability} in a \kl{blind $2$-pebble transducer}.}

\end{figure}

%% file: parts/algo-blind-minimization-normal.tex
\begin{algorithm}[h!]
\SetKwProg{Fn}{Submachine}{}{}
\SetKwProg{Sb}{Submachine}{}{}
\SetKw{In}{in}
\SetKw{To}{to}
\SetKw{Exe}{Simulate}
\SetKw{Out}{Output}
\SetKw{Drop}{Drop}
\SetKw{Erase}{Erase}
\SetKw{Call}{Call}

 \Fn{$\intro*\normalK{\trans}(u)$}{

  	$\rho \defined $ accepting run of $\trans$ over $\lmark u \rmark$;
 	$\lambda \defined $ output function of $\trans$;
				
	\For{$(q,i) \in \rho$}{
	
		\eIf{\normalfont $\trans$ is a leaf of $\utrans$}{
			
			\Out $ \lambda(q, (\lmark u \rmark)[i])$; \tcc{$\trans$ has output in $B^*$;}
		
		}{
	
			\For{$\btrans' \in \lambda(q, (\lmark u \rmark)[i])$}{
				
				$\trans' \defined$ \kl(LAST){head} of $\btrans'$;
		
				\Call $\normalK{\trans'}(u)$; \tcc{$\trans$ makes recursive calls;}
			
			}
		}
		
	}
	
}

 \caption{\label{algo:blind-minimization-original} Submachines
 that behave as the original ones}
\end{algorithm}

%% file: parts/algo-blind-minimization-simplify.tex
\begin{algorithm}[h!]
\SetKw{KwVar}{Variables:}
\SetKwProg{Fn}{Submachine}{}{}
\SetKwProg{Sb}{Submachine}{}{}
\SetKw{Inline}{Inline the code of}
\SetKw{To}{to}
\SetKw{Out}{Produce}
\SetKw{Drop}{Drop}
\SetKw{Erase}{Erase}
\SetKw{Call}{Call}
\SetKw{Move}{Move to}

 \Fn{$\intro*\simBlindK{\trans}~(u)$}{
 
 	\tcc{$\trans$ is not a leaf of $\utrans$ (i.e. it makes calls);}
 
  	$\rho \defined$ accepting {run} of $\trans$ over $\lmark u \rmark$;
	$\forest \defined \forK{\mu}(u)$;  $\lambda \defined \text{output fun. of } \trans$;
				
	\For{$(q,i) \in \rho$}{
	
		\For{$\btrans' \in \lambda(q, (\lmark u \rmark)[i])$}{
%
%
%
%
	
	$\trans' \defined$ \kl(LAST){head} of $\btrans'$;

		\uIf{$i \in \FroK{\forest}(\forest)$}{
				
			\tcc{We can inline the call since $|\FroK{\forest}(\forest)|$ is bounded;}
		
			\Inline $\normalK{\trans'}~(u)$ \label{line:fril} 
			\tcc{(see explanations);}
				
		}
		
		\uElseIf{$\btrans'$ \normalfont is a leaf of $\utrans$\label{line:no-2dt-without-inline}}{
			
			\tcc{Then $\btrans' = \trans'$ and we can inline the call because
			the output of $\trans'$ on input $u$ is bounded;}
				
			\Inline $\normalK{\trans'}~(u)$\label{line:2-way-blind}
			\tcc{(see explanations);}
	
		}
		
		\Else{		

			\tcc{It is not possible to inline the call to $\btrans'$, so
			we make a recursive call;}
					
			\Call \normalfont 
			$\simBlindK{{\trans'}}~(u)$;
			
		}

		}
	}	

}

 \caption{\label{algo:blind-minimization-simplify}Submachines
that try to simulate their recursive calls}
\end{algorithm}

%% file: parts/fig-last-pumpable.tex
\begin{figure}[h!]

\centering
\scalebox{0.8}{
\begin{tikzpicture}{scale=1}

	\newcommand{\coulun}{myDeepPurple}
	\newcommand{\coulde}{myDarkBlue}
	\newcommand{\coultr}{myYellow}

	\fill[\coultr] (1.45,5) rectangle (3.55,4.5);
	\fill[\coultr] (5.45,5) rectangle (7.55,4.5);
	\draw (-0.25,5) rectangle (9.25,4.5);
	\node[above] at (2.5,4.5) {$a_1$};
	\node[above] at (6.5,4.5) {$a_2$};
	\node[above] at (0.05,4.5) {$\lmark$};
	\node[above] at (8.95,4.5) {$\rmark$};

	\node[above] at (0.5,3.9) {$m_0$};
	\node[above] at (1.1,3.9) {$e_1$};	
	\node[above] at (1.7,3.9) {$\ell_1$};	
	\node[above] at (2.5,3.9) {$\cro{a_1}$};
	\node[above] at (3.3,3.9) {$r_1$};
	\node[above] at (3.9,3.9) {$e_1$};
	\node[above] at (4.5,3.9) {$m_1$};
	\node[above] at (5.1,3.9) {$e_2$};
	\node[above] at (5.7,3.9) {$\ell_2$};
	\node[above] at (6.5,3.9) {$\mu(a_2)$};
	\node[above] at (7.3,3.9) {$r_2$};
	\node[above] at (7.9,3.9) {$e_2$};
	\node[above] at (8.5,3.9) {$m_2$};
	

	\fill[\coultr] (1.45,2.5) rectangle (3.55,2);
	\fill[\coultr] (5.45,2.5) rectangle (7.55,2);
	\draw (-0.25,2.5) rectangle (9.25,2);
	\fill[fill = \coulun] (2.7,2.25) circle (0.15);
	\node[above] at (2.3,2) {$a_1$};
	\node[above] at (6.5,2) {$a_2$};
	\node[above] at (0.05,2) {$\lmark$};
	\node[above] at (8.95,2) {$\rmark$};
	
	\node[above] at (0.5,1.4) {$m_0$};
	\node[above] at (1.1,1.4) {$e_1$};	
	\node[above] at (1.7,1.4) {$\ell_1$};	
	\node[above] at (2.5,1.35) {$\mu(\marq{a_1})$};
	\node[above] at (3.3,1.4) {$r_1$};
	\node[above] at (3.9,1.4) {$e_1$};
	\node[above] at (4.5,1.4) {$m_1$};
	\node[above] at (5.1,1.4) {$e_2$};
	\node[above] at (5.7,1.4) {$\ell_2$};
	\node[above] at (6.5,1.4) {$\cro{a_2}$};
	\node[above] at (7.3,1.4) {$r_2$};
	\node[above] at (7.9,1.4) {$e_2$};
	\node[above] at (8.5,1.4) {$m_2$};


	\draw[dotted,  thick] (0.2,5.1) -- (0.2,0.2);
	\draw[dotted,  thick] (0.8,5.1) -- (0.8,0.2);
	\draw[dotted,  thick] (1.45,5.1) -- (1.45,0.2);
	
	\draw[dotted,  thick] (2,5.1) -- (2,0.2);
	\draw[dotted,  thick] (3,5.1) -- (3,0.2);
	\draw[dotted,  thick] (3.55,5.1) -- (3.55,0.2);
	\draw[dotted,  thick] (4.2,5.1) -- (4.2,0.2);

	\draw[dotted,  thick] (7,5.1) -- (7,0.2);
	\draw[dotted,  thick] (6,5.1) -- (6,0.2);
	\draw[dotted,  thick] (5.45,5.1) -- (5.45,0.2);
	\draw[dotted,  thick] (4.8,5.1) -- (4.8,0.2);

	\draw[dotted,  thick] (7.55,5.1) -- (7.55,0.2);
	\draw[dotted,  thick] (8.2,5.1) -- (8.2,0.2);
	\draw[dotted,  thick] (8.8,5.1) -- (8.8,0.2);

	\node[above] at (-1.3,3.25) {\scalebox{1.5}{%
	\textcolor{\coulun}{\text{ {\bfseries $\trans_1$} head}}}};
	\draw[-,line width = 2pt,\coulun](0,3.8) -- (8.2,3.8);
	\draw[-,line width = 2pt, \coulun] (8.2,3.8) arc (90:-90:0.2);
	\draw[-,line width = 2pt,\coulun](8.2,3.4) -- (0.8,3.4);	
	\draw[-, line width = 2pt,\coulun] (0.8,3.4) arc (90:270:0.2);
	\draw[-,line width = 2pt,\coulun](0.8,3) -- (2.5,3);	
	\fill[fill = \coulun,even odd rule] (2.5,3) circle (0.08);

	\node[above] at (-1,0.75) {\scalebox{1.5}{%
	\textcolor{\coulde}{$\trans_2$}}};
	\draw[-,line width = 2pt,\coulde](0,1.3) -- (3,1.3);
	\draw[-,line width = 2pt, \coulde] (3,1.3) arc (90:-90:0.2);
	\draw[-,line width = 2pt,\coulde](3,0.9) -- (1,0.9);	
	\draw[-,line width = 2pt,\coulde] (1,0.9) arc (90:270:0.2);
	\draw[-,line width = 2pt,\coulde](1,0.5) -- (6.5,0.5);	
	\fill[fill = \coulde,even odd rule] (6.5,0.5) circle (0.08);
	\node[above] at (6.5,0.5) {\textcolor{\coulde}{\small $\lambda \neq \movi$}};
	\node[above] at (2.9,2.5) {\textcolor{\coulun}{\small Call $\trans_2$}};

        \draw[->,line width = 2pt,\coulun,dashed](2.5,3) to[out= -150, in = 70] (0,1.4);

\end{tikzpicture}}

\caption{\label{fig:last-pumpable} \kl(LAST){Pumpability} in a \kl{last $2$-pebble transducer}.}

\end{figure}
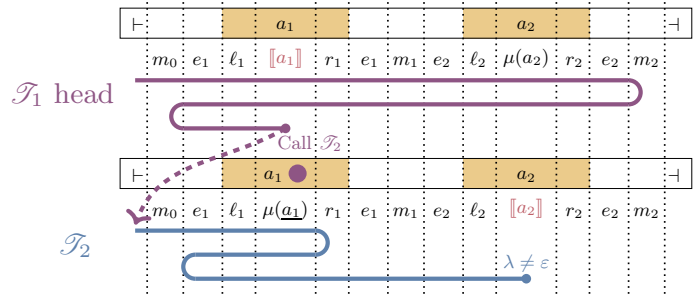

%% file: parts/algo-last-minimization-normal.tex
\begin{algorithm}[h!]
\SetKwProg{Fn}{Submachine}{}{}
\SetKwProg{Sb}{Submachine}{}{}
\SetKw{In}{in}
\SetKw{To}{to}
\SetKw{Exe}{Simulate}
\SetKw{Out}{Output}
\SetKw{Drop}{Drop}
\SetKw{Inline}{Simulate}
\SetKw{Erase}{Erase}
\SetKw{Call}{Call}

 \Fn{$\intro*\normalongK{\trans}{\rho}(v)$}{

  	\tcc{$v \in (A \uplus \marq{A})^*$; $\rho$ is a {run} of $\trans$ over $\lmark v \rmark$;}
 
 	$\lambda \defined $ output function of $\trans$;
				
	\For{$(q,i) \in \rho$}{
	
		\eIf{\normalfont $\trans$ is a leaf of $\utrans$}{
			
			\Out $ \lambda(q, (\lmark v \rmark)[i])$; \tcc{$\trans$ has output in $B^*$;}
		
		}{
	
			\For{$\ltrans' \in \lambda(q, (\lmark v \rmark)[i])$}{
			
				$\trans' \defined$ \kl(LAST){head} of $\ltrans'$;
				$\rho' \defined$ accepting {run} of $\trans'$ on $\lmark \omarK{\tau(v)}{i} \rmark$;

				\Call $\normalongK{\trans'}{\rho'}(\omarK{\tau(v)}{i})$;
				\tcc{Recursive call;}
			
			}
		}
		
	}
	
}

 \Fn{$\intro*\normalongpebK{\trans}{\rho}{i}(v)$}{

  	\tcc{$v \in (A \uplus \marq{A})^*$;
	$\rho$ is a {run} of $\trans$ over $\lmark \omarK{\tau(v)}{i} \rmark$;}
				
	\Inline $\normalongK{\trans}{\rho}~(\omarK{\tau(v)}{i})$;
	
}

 \caption{\label{algo:last-minimization-original} Submachines
 that behave like the original ones}
\end{algorithm}

%% file: parts/fig-slicing.tex
\begin{figure}[h!]

\centering
\begin{tikzpicture}{scale=1}

	\newcommand{\coulun}{myPurple}
	\newcommand{\coulunm}{myDeepPurple}
	\newcommand{\coulde}{myDarkBlue}
	\newcommand{\coultr}{myRed}


	\fill[\coulun] (1,1.5) rectangle (4,2);

	\fill[\coultr] (1,1.5) rectangle (0.5,2);
	\fill[\coultr] (2,1.5) rectangle (2.5,2);
	\fill[\coultr] (7.5,1.5) rectangle (8,2);
	\fill[\coultr] (4,1.5) rectangle (4.5,2);
	
	\draw (0,1.5) rectangle (8.5,2);
	\draw (0,1.5) rectangle (8.5,2);
	\node[above] at (0.25,1.5) {$\lmark$};
	\node[above] at (8.25,1.5) {$\rmark$};

	\node[above] at (2.25,2.1) {$i$};
	\draw[->,  thick] (2.25,2.15) -- (2.25,1.75);

	\node[above] at (4,2.4) {$\textcolor{\coulunm}{\substack{\text{\bfseries positions that}\\
	\text{\bfseries belong to } \ObDoK{i} \smallsetminus  \ObUpK{i} }}$};
	\draw[->,  thick,\coulun] (3.6,2.45) -- (3.6,2.02);

	\node[above] at (7.5,2.4) {$\textcolor{\coultr}{\substack{\text{\bfseries positions that}\\
	\text{\bfseries belong to }  \ObUpK{i} }}$};
	\draw[->,  thick, \coultr] (7.75,2.45) -- (7.75,2.05);


	\draw[dotted,  thick] (0.5,2.1) -- (0.5,0.2);
	\draw[dotted,  thick] (1,2.1) -- (1,0.2);

	\draw[dotted,  thick] (2,2.1) -- (2,0.2);
	\draw[dotted,  thick] (2.5,2.1) -- (2.5,0.2);

	\draw[dotted,  thick] (4,2.1) -- (4,0.2);
	\draw[dotted,  thick] (4.5,2.1) -- (4.5,0.2);
	
	\draw[dotted,  thick] (7.5,2.1) -- (7.5,0.2);
	\draw[dotted,  thick] (8,2.1) -- (8,0.2);

	\draw[-,line width = 2pt,\coulde](0,1.3) -- (0.5,1.3);
	\draw[-,line width = 2pt,\coultr](0.5,1.3) -- (1,1.3);
	\draw[-,line width = 2pt,\coulun](1,1.3) -- (2,1.3);
	\draw[-,line width = 2pt,\coultr](2,1.3) -- (2.5,1.3);
	\draw[-,line width = 2pt,\coulun](2.5,1.3) -- (4,1.3);
	\draw[-,line width = 2pt,\coultr](4,1.3) -- (4.3,1.3);
	\draw[-,line width = 2pt, \coultr] (4.3,1.3) arc (90:-90:0.2);
	\draw[-,line width = 2pt,\coultr](4.3,0.9) -- (4,0.9);	
	\draw[-,line width = 2pt,\coulun](4,0.9) -- (2.5,0.9);
	\draw[-,line width = 2pt,\coultr](2.5,0.9) -- (2,0.9);
	\draw[-,line width = 2pt,\coulun](2,0.9) -- (1,0.9);
	\draw[-,line width = 2pt,\coultr](1,0.9) -- (0.7,0.9);		
	\draw[-,line width = 2pt,\coultr] (0.7,0.9) arc (90:270:0.2);
	\draw[-,line width = 2pt,\coultr](0.7,0.5) -- (1,0.5);	
	\draw[-,line width = 2pt,\coulun](1,0.5) -- (2,0.5);	
	\draw[-,line width = 2pt,\coultr](2,0.5) -- (2.5,0.5);
	\draw[-,line width = 2pt,\coulun](2.5,0.5) -- (4,0.5);	
	\draw[-,line width = 2pt,\coultr](4,0.5) -- (4.5,0.5);	
	\draw[-,line width = 2pt,\coulde](4.5,0.5) -- (7.5,0.5);
	\draw[-,line width = 2pt,\coultr](7.5,0.5) -- (8,0.5);
	\draw[-,line width = 2pt,\coulde](8,0.5) -- (8.5,0.5);

	\node[above] at (0.25,0.9) {\textcolor{\coulde}{$\rho_1$}};
	\node[above] at (0.75,0.9) {\textcolor{\coultr}{$\rho_2$}};
	\node[above] at (1.5,0.9) {\textcolor{\coulunm}{$\rho_3$}};
	\node[above] at (2.25,0.9) {\textcolor{\coultr}{$\rho_4$}};
	\node[above] at (3.25,0.9) {\textcolor{\coulunm}{$\rho_5$}};
	\node[above] at (4.25,0.9) {\textcolor{\coultr}{$\rho_6$}};
	
	\node[above] at (3.25,0.5) {\textcolor{\coulunm}{$\rho_7$}};
	\node[above] at (2.25,0.5) {\textcolor{\coultr}{$\rho_8$}};
	\node[above] at (1.5,0.5) {\textcolor{\coulunm}{$\rho_9$}};
	\node[above] at (0.8,0.5) {\textcolor{\coultr}{$\rho_{10}$}};
	
	\node[above] at (1.5,0.1) {\textcolor{\coulunm}{$\rho_{11}$}};
	\node[above] at (2.25,0.1) {\textcolor{\coultr}{$\rho_{12}$}};
	\node[above] at (3.25,0.1) {\textcolor{\coulunm}{$\rho_{13}$}};
	\node[above] at (4.25,0.1) {\textcolor{\coultr}{$\rho_{14}$}};

	\node[above] at (6,0.1) {\textcolor{\coulde}{$\rho_{15}$}};;
	\node[above] at (7.75,0.1) {\textcolor{\coultr}{$\rho_{16}$}};
	\node[above] at (8.25,0.1) {\textcolor{\coulde}{$\rho_{17}$}};

\end{tikzpicture}

\caption{\label{fig:slicing} \kl{Slicing} of a run $\rho$ with respect to $i$ and $\forest$.}

\end{figure}

%% file: parts/algo-last-minimization-simplify.tex
\begin{algorithm}[h!]
\SetKw{KwVar}{Variables:}
\SetKwProg{Fn}{Submachine}{}{}
\SetKwProg{Sb}{Submachine}{}{}
\SetKw{In}{in}
\SetKw{To}{to}
\SetKw{Inline}{Inline the code of}
\SetKw{Out}{Produce}
\SetKw{Drop}{Drop}
\SetKw{Erase}{Erase}
\SetKw{Call}{Call}
\SetKw{Move}{Move to}

 \Fn{$\intro*\simLastK{\trans}{\rho}~(v)$}{

	\tcc{$\trans$ is not a leaf of $\utrans$ (i.e. it makes calls);}
	 
  	\tcc{$v \in (A \uplus \marq{A})^*$;
	$\rho$ is a run of $\trans$ over $\lmark v \rmark$;}
	
	$u \defined \tau(v)$; $\forest \defined \forK{\mu}(u)$; 
	$\lambda \defined \text{output function of } \trans$;

	\For{$(q,i) \in \rho$}{
	
		\For{$\ltrans' \in \lambda(q, (\lmark v \rmark)[i])$}{

	$\trans' \defined$ \kl(LAST){head} of $\ltrans'$;	
	$\rho' \defined \text{accepting run of $\trans'$ over $\lmark \omarK{u}{i} \rmark$}$;

%
%
%
%
		\label{line:N-slice}$\rho'_1, \cdots, \rho'_N \defined$ \kl{slicing} %
		of $\rho'$ with respect to $\forest$ and $i$;
	
		\For{$j=1$ \To $N$}{
		
			$(q_1,i_1) \runs{} \cdots (q_n,i_n) \defined \rho'_j$
		
			\uIf{$i_1, \dots, i_n \in \ObUpK{i}$}{
			
				\tcc{We inline the call because $n$ is bounded;}	
	
				\Inline $\normalongK{\trans'}{\rho'_j}~(\omarK{u}{i})$;			
			
			}
			\uElseIf{$i_1, \dots, i_n \in \ObDoK{i}$}{
			
				\tcc{We can inline the call because the positions 
				$i_1, \dots, i_n$ are ``below'' $i$ in $\forest$;}
				
				\Inline $\normalongK{\trans'}{\rho'_j}~(\omarK{u}{i})$;
		
			}
			\uElseIf{\label{line:testtest-no2DT}$\ltrans'$ {\normalfont is a leaf of $\utrans$}}{
				
				\tcc{\label{line:key-blind-minimization}%
				The output of $\ltrans' = \trans'$
				along $\rho'_j$ is empty;}

			}
			
			\Else{

				\tcc{It is not possible to inline the call to $\ltrans'$, so
				we make a recursive call;}
					
				\Call \normalfont 
				$\simLastK{{\trans'}}{\rho'_j}~(\omarK{u}{i})$;
								
			}

		}
	
 		}
	}	

}

 \caption{\label{algo:last-minimization-simplify}Submachines
that try to simulate their recursive calls}

\end{algorithm}

%% file: parts/appendix.tex
\section{Omitted proofs of \cref{sec:blind-proof}}

\subsection{Proof of \cref{lem:growth-blind}}

Let $w_i \in A^*$ be such that $\mu(w_i) = \ell_i$,
 $w'_i \in A^*$ be such that $\mu(w'_i) = r_i$ and
$u_i \defined w_i a_i w'_i$, for $1 \le i \le k$.
Let $v_0, \dots, v_k \in A^*$ such that
$\mu(v_i) = m_i$ for all $1 \le i \le k$.
Let $f_i$ be the function
computed by $\trans_i$ for all $1 \le i \le k$.

It is easy to see that $|f_k(v_0 u_1^{X} \cdots u_k^{X} v_k)| \ge (X{-}2)$.
Indeed, $\trans_k$ must produce at least one letter
when reading the letter $a_{\sigma(k)}$ of each factor
$u_{\sigma(k)}$ (with possibly an exception
for the borders, hence the ${-}2$). We then show
by induction on $k \ge i \ge 1$
that $|f_i(v_0 u_1^{X} \cdots u_k^{X} v_k)| \ge (X{-}2)^{k-i+1}$.
The upper bound follows since $\btrans$ is
a \kl{blind $k$-pebble transducer}, thus
$|f(v_0 u_1^{X} \cdots u_k^{X} v_k)| = \bigT(X^k)$.

\subsection{Additional arguments in \cref{ssec:minimize-blind}}

We first justify that the construction of  $\ov{\utrans}$ in \cref{ssec:minimize-blind} indeed
reduces the recursion height of $\utrans$ by $1$. This statement was claimed
on \cpageref{sta:k:blind}.

\begin{claim} The machine $\ov{\utrans}$ described in  \cref{ssec:minimize-blind}
has recursion height $k{-}1$.
\end{claim}

\begin{proof} Recall that the recursion height corresponds to the
number of nested \textbf{Call} instructions, plus $1$ (for the \kl(BLIND){submachine}
which is the \kl(BLIND){head}).
We first show by (decreasing) induction on $1 \le \ell \le k$
that if  $\trans$ is the \kl(BLIND){head} of a subtree of $\utrans$ whose
recursion height is $1 \le \ell \le k$, then
$\normalK{\trans}$ has recursion height
$\ell$ as well.\\
Then, we show by (decreasing) induction on $1 \le \ell \le k$
that if  $\trans$ is the \kl(BLIND){head} of a subtree of $\utrans$ whose
recursion height is $\ell$, then $\simBlindK{\trans}$ has recursion
height $\ell{-}1$. Indeed, the base case $\ell = 2$ is justified by
\cref{line:2-way-blind} in \cref{algo:blind-minimization-simplify}
(there are no calls since we inline all the computations).
For $\ell > 2$ the function $\simBlindK{\trans}$ inlines 
$\normalK{\trans'}$ of recursion height $\ell{-}1$ and
makes a recursive call to $\simBlindK{\trans'}$ whose
height is $\ell{-}2$  by induction hypothesis.\\
The result follows since the \kl(BLIND){head} of $\utrans$
has  recursion height $k$ by definition of a \kl{blind $k$-pebble transducer}.
\end{proof}

Since we have justified in the main paper how each function of $\ov{\utrans}$ 
can be implemented by a \kl{two-way transducer}, then $\ov{\utrans}$ is indeed a
\kl{blind $(k{-}1)$-pebble transducer}. Now, we justify a claim 
of \cpageref{sta:k:2}, that is used to show that the output in \cref{line:2-way-blind}
of  \cref{algo:blind-minimization-simplify} must be bounded.

\begin{claim} Assume that, in the execution of $\ov{\utrans}$ on input $u \in A^*$,
we reach \cref{line:2-way-blind}  in  \cref{algo:blind-minimization-simplify} and that
$\valK{\trans'}{u}(i') \neq \movi$ for $i' \not \in \FroK{\forest}(\forest)$.
Then the conditions of \cref{lem:k-nodes-blind-key} hold, that is
there exists $\forest \in \FacK{\mu}{}(u)$, a sequence $\trans_1, \dots, \trans_k$
of \kl(BLIND){submachines} of $\utrans$ and a sequence of positions 
$1 \le i_1, \dots, i_k \le |u|$ such that:
\begin{itemize}
\item $\trans_1$ is the \kl(BLIND){head} of $\utrans$;
\item  for all $1 \le j \le k{-}1$, $|\valK{\trans_j}{u}(i_j)|_{\trans_{j+1}} \neq 0$
and $\valK{\trans_k}{u}(i_k) \neq \movi$;
\item for all $1 \le j \le k$, $i_j \not \in \FroK{\forest}(\forest)$
(i.e. $\originK{\forest}(i_j) \in \IteK{\forest}$).
\end{itemize}
\end{claim}

\begin{proof} Let $\trans$ be a \kl(BLIND){submachine} of $\utrans$
which is not a leaf (i.e. it labels the \kl(BLIND){head} of a subtree of height $>1$).
We claim that for $\simBlindK{\trans}(u)$ to be called when executing
$\ov{\utrans}$ on input $u \in A^*$, there must exist a sequence $\trans_1, \dots, \trans_{\ell}$
of \kl(BLIND){submachines} of $\utrans$ and a sequence of positions 
$1 \le i_1, \dots, i_{\ell-1} \le |u|$ such that:
\begin{itemize}
\item $\trans_1$ is the \kl(BLIND){head} of $\utrans$ and $\trans_{\ell} = \trans$;
\item  for all $1 \le j \le \ell{-}1$, $|\valK{\trans_j}{u}(i_j)|_{\trans_{j+1}} \neq 0$;
\item for all $1 \le j \le \ell{-}1$, $i_j \not \in \FroK{\forest}(\forest)$
where $\forest \defined \forK{\mu}(u)$.
\end{itemize}
This result can be checked by induction. Intuitively, it means that to systematically
avoid inlinings, we have to make recursive calls in a sequence of
positions which are never in the \kl{frontier} on the root.\\
Finally, by considering $\trans$ the \kl(BLIND){head} of a subtree of height $2$, we see that the conditions
of  \cref{lem:k-nodes-blind-key}  must hold if we reach
we reach \cref{line:2-way-blind}  in $\simBlindK{\trans}$ and if
$\valK{\trans'}{u}(i') \neq \movi$ for $i' \not \in \FroK{\forest}(\forest)$.
\end{proof}

To conclude about the omitted proofs in \cref{ssec:minimize-blind},
it remains to show \cref{lem:k-nodes-blind-key}.
This is the purpose of \cref{ssec:proof:lem-blind}.

\subsection{Proof of \cref{lem:k-nodes-blind-key}}

\label{ssec:proof:lem-blind}

Assume that the conditions of \cref{lem:k-nodes-blind-key} hold
and let $\nod_j \defined \originK{\forest}(i_j)$ for all $1 \le j \le k$.
If the $\nod_j$ are pairwise \kl(NOD){independent},
then each $\nod_j$ is surrounded by two nodes
whose \kl{frontiers} cannot contain a position $i_{j'}$
for some $1 \le j' \le k$. The image of the factor of $u$
which is below these nodes provides an idempotent $e_j$.
It can easily be concluded that $\utrans$ is \kl(BLIND){pumpable}
(see also \cite[Lemma E.5]{doueneau2022hiding}).

\input{parts/fig-copy-forest.tex}

Now, we suppose that the $\nod_j$ are
not necessarily pairwise \kl(NOD){independent}.
Let us show how to make the number of \kl(NOD){dependent} couples
of $(\nod_{j_1}, \nod_{j_2})$ decrease strictly, while preserving the properties of \cref{lem:k-nodes-blind-key}.
Indeed, repeating this process will enable us to
make all the nodes pairwise \kl(NOD){independent}.
Assume that $\nod_{\ell_1}$ \kl(NOD){observes} $\nod_{\ell_2}$
for some $1 \le \ell_1 \neq \ell_2 \le k$. To simplify the proof,
we assume that $\nod_{\ell_2}$ is an ancestor
of $\nod_{\ell_1}$ (the case of the immediate sibling of
an ancestor is similar).
Let $\forest'$ be $\forest$ in which the subtree
 $\nod_{\ell_2}$ has been copied $3$ times
(since $\nod_{\ell_2}$ is an \kl{iterable node},
then $\forest'$ still a \kl{$\mu$-forest}), see \cref{fig:making-nodes-far}.
We define for $1 \le j \le k$ the nodes $\nod'_{j} \in \NodK{\forest'}$
as follows:
\begin{itemize}
\item if $j = \ell_2$, then $\nod'_{j}$ is (the root of) the third copy
of $\nod_j$;
\item else if $j$ is such that $\nod_j$ was a descendant
of $\nod_{\ell_2}$ (including $\nod_{\ell_1}$), then we let $\nod'_{j}$
be the corresponding node in the first copy of $\nod_j$;
\item else $\nod_j$ was in the rest of $\forest$,
and we let  $\nod'_j$ be the corresponding
node in $\forest'$.
\end{itemize}

Observe that now, $\nod'_{\ell_1}$ and $\nod'_{\ell_2}$
are not \kl(NOD){dependent}. Furthermore if $\nod_{j_1}$
and $\nod_{j_2}$ were \kl(NOD){independent}, then 
$\nod'_{j_1}$ and $\nod'_{j_2}$ are also \kl(NOD){independent}.
Let $u' \in A^*$ be the word such that $\forest' \in \FacK{\mu}{}(u')$.
We also define $1 \le i'_1, \dots, i'_k \le |u'|$ as the positions
which correspond to the former $1 \le i_1, \dots, i_k \le |u|$
in the \kl{frontiers} of $\nod_1', \dots, \nod'_k$
in the new \kl{$\mu$-forest} $\forest'$. The conditions of \cref{lem:k-nodes-blind-key}
still hold, because $\valK{\trans_j}{u}(i_j) = \valK{\trans_j}{u'}(i'_j)$
(indeed, we have only duplicated an \kl{iterable node}, which does
neither modify the \kl{context} around $i'_j$ nor its \kl{crossing sequence}).

\section{Omitted proofs of \cref{sec:last-proof}}

\subsection{Proof of \cref{lem:growth-last}}

The proof is similar to that of \cref{lem:growth-blind}.
Let $w_i \in A^*$ be such that $\mu(w_i) = \ell_i$,
 $w'_i \in A^*$ be such that $\mu(w'_i) = r_i$,
$u_i \defined w_i a_i w'_i$ and $\marq{u_i} \defined w_i \marq{a_i} w'_i$,
for $1 \le i \le k$.
Let $v_0, \dots, v_k \in A^*$ such that
$\mu(v_i) = m_i$ for all $1 \le i \le k$.
Let $f_i$ be the function
computed by $\trans_i$ for all $1 \le i \le k$.

To simply the proof, we assume that $\sigma : [1{:}k] \fonc [1{:}k]$
is the identity function. We then observe that for all $X \ge 2$,
for all $1 \le Y \le X{-}2$:
\begin{equalign*}
f_k(v_0 u_1^X \cdots v_{k-2} (u_{k-1}^Y \marq{u_{k-1}} u_{k-1}^{X-Y-1}) v_{k-1} u_k^X) \ge (X-2).
\end{equalign*}
Observe that the use of $u_{k-1}^Y \marq{u_{k-1}}  u_{k-1}^{X-Y-1}$ means
that the result holds independently from the factor in which the call (i.e. the mark)
to $\trans_k$ was done. Finally, we conclude by induction
in a similar way to \cref{lem:growth-blind}.

\subsection{Additional arguments in \cref{ssec:minimize-last}}

We first justify that the construction of  $\ov{\utrans}$ in \cref{ssec:minimize-blind} indeed
reduces the recursion height of $\utrans$ by $1$. This statement was claimed
on \cpageref{sta:last:k}.

\begin{claim} The machine $\ov{\utrans}$ described in  \cref{ssec:minimize-last}
has recursion height $k{-}1$.
\end{claim}

\begin{proof} Recall that the recursion height corresponds to the
number of nested \textbf{Call} instructions, plus $1$ (due to the \kl(LAST){head}).
We first show by (decreasing) induction on $1 \le \ell \le k$
that if  $\trans$ is the \kl(LAST){head} of a subtree of $\utrans$ whose
recursion height is $1 \le \ell \le k$, then
$\normalongK{\trans}{\rho}$ has recursion height
$\ell$ as well.\\
Then, we show by (decreasing) induction on $2 \le \ell \le k$
that if  $\trans$ is the \kl(LAST){head} of a subtree of $\utrans$ whose
recursion height is $\ell$, then $\simLastK{\trans}{\rho}$ has recursion
height $\ell{-}1$. Indeed, the base case $\ell = 2$ is justified by
\cref{line:key-blind-minimization} in \cref{algo:last-minimization-simplify}
(there are no calls).
For $\ell > 2$ the function $\simLastK{\trans}{\rho}$ inlines some
$\normalongK{\trans'}{\rho'_j}$ of recursion height $\ell{-}1$ and
makes recursive calls to $\simLastK{\trans'}{\rho_j}$ whose
height is $\ell{-}2$  by induction hypothesis.\\
The result follows since the \kl(LAST){head} of $\utrans$
has  recursion height $k$ by definition of a \kl{last $k$-pebble transducer}.
\end{proof}

Now, let us justify a claim  of \cpageref{sta:last:2}, that is used to show
that the output of \cref{line:key-blind-minimization}
in \cref{algo:last-minimization-simplify} is indeed empty.

\begin{claim} Assume that, in the execution of $\ov{\utrans}$ on input $u \in A^*$,
we reach \cref{line:key-blind-minimization}
in \cref{algo:last-minimization-simplify} and that
the output of $\trans'$ along $\rho'_j$ is not empty.
Then the conditions of \cref{lem:k-nodes-last-key} hold, that is
there exists $\forest \in \FacK{\mu}{}(u)$, a sequence $\trans_1, \dots, \trans_k$
of \kl(LAST){submachines} of $\utrans$ and a sequence of positions 
$1 \le i_1, \dots, i_k \le |u|$ such that:
\begin{itemize}
\item $\trans_1$ is the \kl(LAST){head} of $\utrans$;
\item $|\valK{\trans_1}{u}(i_1)|_{\trans_2} \neq 0$ and
$\valK{\trans_k}{\omarK{u}{i_{k-1}}}(i_k) \neq \movi$;
\item for all $2 \le j \le k{-}1$, $| \valK{\trans_j}{\omarK{u}{i_{j-1}}}(i_j) |_{\trans_{j+1}} \neq 0$;
\item for all $1 \le j \le k{-}1$, $\originK{\forest}(i_j)$
and $\originK{\forest}(i_{j+1})$ are \kl(NOD){independent};
\end{itemize}
\end{claim}

\begin{proof} Let $\trans$ be a \kl(LAST){submachine} of $\utrans$
which is neither a leaf (i.e. it is the \kl(LAST){head} of a subtree of height $>1$) nor the
\kl(LAST){head} of $\utrans$ (i.e. not the root of $\utrans$).
We claim that for $\simLastK{\trans}{\rho}~(v)$ to be called whithin
the execution of $\ov{\utrans}$ on input $u \in A^*$, there must exist a sequence
$\trans_1, \dots, \trans_{\ell}$
of \kl(LAST){submachines} of $\utrans$ and a sequence of positions 
$1 \le i_1, \dots, i_{\ell-1} \le |u|$ such that:
\begin{itemize}
\item $\trans_1$ is the \kl(LAST){head} of $\utrans$;
\item $v = \omarK{u}{i_{\ell-1}}$;
\item $|\valK{\trans_1}{u}(i_1)|_{\trans_2} \neq 0$ and
for all $2 \le j \le k{-}1$, $| \valK{\trans_j}{\omarK{u}{i_{j-1}}}(i_j) |_{\trans_{j+1}} \neq 0$;
\item for all $1 \le j \le \ell{-}2$, $\originK{\forest}(i_j)$
and $\originK{\forest}(i_{j+1})$ are \kl(NOD){independent},
where we define $\forest \defined \forK{\mu}(u)$;
\item for all $(i,q) \in \rho$, $\originK{\forest}(i_{\ell-1})$ and $\originK{\forest}(i)$
are \kl(NOD){independent}
\end{itemize}
This result can be checked by induction. Intuitively, the two crucial last point follows
from the fact that we only make recursive calls in portions of runs
whose positions are not \kl(POS){dependent} on the calling position.\\
Finally, by considering $\trans$ the \kl(LAST){head} of a subtree of height $2$, we see that the conditions
of  \cref{lem:k-nodes-last-key}  must hold if we reach
\cref{line:key-blind-minimization} in some $\simLastK{\trans}{\rho}$
called in $\ov{\utrans}$ and if
the output of $\trans'$ along $\rho'_j$ is not empty.
\end{proof}

To conclude about the omitted proofs in \cref{ssec:minimize-last},
it remains to show \cref{lem:k-nodes-last-key,lem:struct-O2}.
This is the purpose of \cref{ssec:proof:lem-last,ssec:proof:lem-O2}.

\subsection{Proof of \cref{lem:struct-O2}}

\label{ssec:proof:lem-O2}

Let $1 \le i \le |u|$, $\nod \defined \originK{\forest}(i)$ and $\nod_{1}$ (resp. $\nod_2$)
be its immediate left (resp. right) sibling (they exist whenever $\nod \in \IteK{\forest}$,
i.e. here $\nod \neq \forest$). We show that:
\begin{equalign*}
\ObDoK{i} \smallsetminus \ObUpK{i}  = [\min(\FroK{\forest}(\nod_1)) : \max(\FroK{\forest}(\nod_2))] \smallsetminus \{\FroK{\forest}(\nod_1), \FroK{\forest}(\nod),
\FroK{\forest}(\nod_2)\}.
\end{equalign*}

Let us assume that $\nod_1$ and $\nod_2$ are \kl{iterable nodes} of $\forest$
(the other cases are similar). By considering the forest of \cref{fig:observation},
it can be noted that $\ObDoK{i}$ is the interval
$[\min(\FroK{\forest}(\nod_1)) : \max(\FroK{\forest}(\nod_2))]$. We conclude
since $\nod, \nod_1$ and $\nod_2$ are the only \kl{iterable nodes}
that both \kl(NOD){observe} $\nod$ and that $\nod$ \kl(NOD){observes}.

Therefore $\ObDoK{i} \smallsetminus \ObUpK{i}$ is the union of a bounded
number of intervals (since the \kl{frontiers} have bounded size).
It is easy to observe that the ``borders'' of these intervals
can easily be recovered by a \kl{lookaround}, if $\nod$ (or $i$) is given.

\subsection{Proof of \cref{lem:k-nodes-last-key}}

\label{ssec:proof:lem-last}

The proof is similar to that of \cref{lem:k-nodes-blind-key}.
The goal is to show that the for $1 \le j \le k$, the
$\nod_j \defined \originK{\forest}(i_{j})$ 
can be chosen pairwise \kl(NOD){independent} 
(in the hypothesis, it is only assumed for the consecutive pairs
$(\nod_j, \nod_{j+1})$).

For this, we show once more how to make the
number of \kl(NOD){dependent} nodes decrease strictly,
while preserving the properties of \cref{lem:k-nodes-last-key}.
Assume that $\nod_{\ell_1}$ \kl(NOD){observes} $\nod_{\ell_2}$
for some $1 \le \ell_1 \neq \ell_2 \le k$ (note that $\ell_1$
and $\ell_2$ are not consecutive). To simplify the proof,
we assume that $\nod_{\ell_2}$ is an ancestor
of $\nod_{\ell_1}$ (the case of the immediate sibling of
an ancestor is similar). We build $\forest' \in \FacK{\mu}{}(u')$
as in the proof of \cref{lem:k-nodes-blind-key} (see \cref{fig:making-nodes-far}),
and define the new nodes $\nod'_{1}, \dots, \nod'_{k} \in \NodK{\forest'}$
in the same way. We also define $1 \le i'_1, \dots, i'_k \le |u'|$ as the positions
which correspond to the former $1 \le i_1, \dots, i_k \le |u|$
adapted to the new nodes $\nod'_{1}, \dots, \nod'_{k}$.

Now, we justify that $\valK{\trans_j}{\omarK{u}{i_{j-1}}}(i_j) = \valK{\trans_j}{\omarK{u'}{i'_{j-1}}}(i'_j) $
for all $2 \le j \le k$. This is the only difference with the proof of
of \cref{lem:k-nodes-blind-key} that we need to treat:
\begin{itemize}
\item if both $i_{j-1}$ and $i_{j}$ belong to the subtree
rooted in $\nod_{\ell_2}$, then $j \neq \ell_2$  (since otherwise $i_{j}$
and $i_{j-1}$ would be \kl(NOD){dependent})
and similarly $j{-}1 \neq \ell_2$.
The result holds because we only iterate an \kl{iterable node};
\item if both $i_{j-1}$ and $i_{j}$ do not belong to this subtree,
the argument is similar;
\item if $i_{j-1}$ is in the subtree but not $i_j$ (the converse is similar),
we use the fact that these two nodes are \kl(NOD){independent}.
Indeed, it implies that $i_j$ cannot be ``below'' an immediate sibling of
$\nod_{\ell_2}$. Hence duplicating this \kl{iterable node} will not
change the monoid value between positions $i_{j-1}$ and $i_j$. 
\end{itemize}

\section{Proof of \cref{the:fail-last-last}}

Let $A$ be an alphabet and $k \ge 1$.
We define the tree language  $T^{k}_{A}$ as the
set of trees such that all root-to-leaf
branches have exactly $k$ nodes
(hence the tree has height $k$),
whose leaves are labelled by words of $A^*$
and whose inner nodes have no labels.
As observed for \kl{factorization forests},
$T^k_A$ can be seen as a regular word language
over the alphabet $\apar{A} \defined A \uplus \{\lefttree, \righttree\}$.
Given such a tree, we say that the
root has \intro{height} $1$, its children \kl{height} $2$,
etc. and the leaves
have \kl{height} $k$.

Now, we describe for all $k \ge 1$ a
function $\intro*\zebraK{k} : T^{k+1}_A \fonc T^{2k+1}_{A\uplus \#}$
which goes from  words  to words (i.e. it works on the word representation
of the trees).
This function was introduced by Boja{\'{n}}czyk in \cite{bojanczyk2022growth}.
Intuitively, it produces a tree whose leaves
labels are tuples $u \# v$ for $u,v$ labels of the original tree,
but the ordering of these tuples is very specific.

Let us first describe the function $\zebraK{1}$.
It takes as input a tree of height $2$ of shape
$\tree{\tree{u_1},\tree{u_2}, \cdots, \tree{u_n}}$ and it produces
a tree of height  $3$ whose $i$-th child of the root
is $\tree{\tree{u_i\#u_1}, \cdots  ,\tree{u_i\#u_n}}$
(i.e. the tuples are ordered lexicographically).
This function be implemented by a \kl{$3$-pebble transducer}
(i.e.  a \kl{last-last $3$-pebble transducer}) which uses its two first pebble
to see which leaves have to be produced, 
and  the last layer to indeed output these leaves.
Observe that $\zebraK{1}$ can be seen as a variant
of the function $\ulsq$ presented in \cref{ex:ulsq}.

Now, the function $\zebraK{2}$ is described formally
in \cref{algo:zebra}.

\input{parts/algo-zebra.tex}

Observe that $\zebraK{2}$ no longer produces
the tuples $u \# v$ in a lexicographic
ordering. Indeed, it corresponds to two ranges over the leaves
of the original tree (one with $i_{1},i_2$ and one with $j_1,j_2)$
which are highly entangled. It is easy to guess how to extend
 \cref{algo:zebra}, in order to define $\zebraK{k}$ using $2k$ nested loops.
 
 Originally, the  $\zebraK{k}$ functions were used in order
 to show that the minimal number of layers and the growth of the output
 do not coincide for \kl{pebble transducers}, as claimed
 in \cref{the:fail-pebble}.

\begin{theorem}[{\cite[section 3]{bojanczyk2022growth}}]
\label{the:fail-pebble}
For all $k \ge 1$, the function $\zebraK{k}$ is such that
$|\zebraK{k}(u)| = \bigO(|u|^2)$. Furthermore, it
can be computed by a \kl{$(2k{+}1)$-pebble transducer}
but not by a \kl{$2k$-pebble transducer}.
\end{theorem}

As a consequence, $\zebraK{k}$ cannot be computed by
a \kl{last-last $2k$-pebble transducer}. To show \cref{the:fail-last-last},
it is thus sufficient for us to justify that  $\zebraK{k}$ can be computed
by \kl{last-last $(2k{+}1)$-pebble transducer}. This is indeed
the case: it uses its $2k$ first layers to describe the nested loops on $i_1, j_1, i_2,j_2, \dots, i_k, j_k$
and the last one to range over the labels of the tuple of leaves and output them
(see \cref{algo:zebra}). The key observation is that it only needs to see the two last
loop indexes, since this information is sufficient to find their children.

%% file: parts/fig-copy-forest.tex
\begin{figure}[h!]

\centering
	\newcommand{\coulun}{myDeepPurple}
	\newcommand{\coulde}{myRed}
	\newcommand{\coultr}{myDarkBlue}

	\begin{subfigure}[b]{0.4\textwidth}
	\centering
        	
         	\begin{tikzpicture}{scale=1}
              	
            	\draw[-,line width = 2pt,\coulun](0,0) -- (-0.5,-2) -- (0.5,-2) -- cycle;
            	
            	\node[above] at (0.4,-0.5)  {\textcolor{\coultr}{$\nod_{\ell_2}$}};
            	\node[above] at (0.2,-1.8) {\textcolor{\coulun}{$\nod_{\ell_1}$}};
            
            	\draw[double, very thick] (-2,0.5) -- (2,0.5);
            	\draw[very thick] (-1.2,0.5) -- (-1.2,0);
            	\draw[very thick] (1.2,0.5) -- (1.2,0);
            	\draw[very thick] (0,0.5) -- (0,0);

            	\fill[fill = \coultr,even odd rule] (0,0) circle (0.15);

            	\fill[fill = \coulun,even odd rule] (0,-1.2) circle (0.15);

                \end{tikzpicture}
            
	\caption{Original \kl{$\mu$-forest} $\forest$}
	\end{subfigure}
	\begin{subfigure}[b]{0.56\textwidth}
	\centering
        	
         	\begin{tikzpicture}{scale=1}
              	
            	\draw[-,line width = 2pt,\coulun](0,0) -- (-0.5,-2) -- (0.5,-2) -- cycle;
            	
            	\node[above] at (1.6,-0.5)  {\textcolor{\coultr}{$\nod'_{\ell_2}$}};
            	\node[above] at (-1,-1.85) {\textcolor{\coulun}{$\nod'_{\ell_1}$}};
            
            	\draw[double, very thick] (-3.2,0.5) -- (3.2,0.5);
            	\draw[very thick] (-2.4,0.5) -- (-2.4,0);
            	\draw[very thick] (-1.2,0.5) -- (-1.2,0);
            	\draw[very thick] (1.2,0.5) -- (1.2,0);
            	\draw[very thick] (2.4,0.5) -- (2.4,0);
            	\draw[very thick] (0,0.5) -- (0,0);

            	\draw[-,line width = 2pt,\coulun](1.2,0) -- (0.7,-2) -- (1.7,-2) -- cycle;
            
            	\draw[-,line width = 2pt,\coulun](-1.2,0) -- (-0.7,-2) -- (-1.7,-2) -- cycle;

            	\fill[fill = \coultr,even odd rule] (1.2,0) circle (0.15);

            	\fill[fill = \coulun,even odd rule] (-1.2,-1.2) circle (0.15);

                \end{tikzpicture}
            
	\caption{Modified \kl{$\mu$-forest} $\forest'$}
	\end{subfigure}

\caption{\label{fig:making-nodes-far} Duplicating
a subtree in $\forest$ so that $\nod'_{\ell_1}$
and $\nod'_{\ell_2}$ are not \kl(NOD){dependent}.}

\end{figure}

%% file: parts/algo-zebra.tex
\begin{algorithm}[h!]
\SetKwProg{Fn}{Function}{}{}
\SetKwProg{Sb}{Submachine}{}{}
\SetKw{In}{in}
\SetKw{To}{to}
\SetKw{Exe}{Simulate}
\SetKw{Out}{Output}
\SetKw{Drop}{Drop}
\SetKw{Erase}{Erase}
\SetKw{Call}{Call}
\SetKw{Range}{ranging from left to right on the children of}

 \Fn{$\zebraK{2}(u)$}{

  	$u \in A^*$ represents a tree of \kl{depth} $k{+}1$; 
	
	$i_0 \defined j_0 \defined$ the root of $u$;
				
	\For{$i_1$ \Range $i_0$}{
	
		\Out $\lefttree$;
	
		\For{$j_1$ \Range $j_0$}{
			
			\Out $\lefttree$;
			
			\For{$i_2$ \Range $i_1$}{
				
				\Out $\lefttree$;
			
				\For{$j_2$ \Range $j_1$}{
				
					 \tcc{Depth $3$: $i_2$ and $j_2$ are leaves;}
					
					$u \defined$ label of $i_2$; $v\defined$ label of $j_2$;
				
					\Out  $\lefttree u \# v \righttree$
			
				}
			
				\Out $\righttree$;
			
			}
			
			\Out $\righttree$;
		
		}
		
		\Out $\righttree$;
	
	}

}

 \caption{\label{algo:zebra} Computing
 the $\zebraK{2}$ function}
\end{algorithm}